\documentclass[pdflatex,sn-nature]{sn-jnl}%

\usepackage{graphicx}%
\usepackage{multirow}%
\usepackage{gensymb}%
\usepackage{amsmath,amssymb,amsfonts}%
\usepackage{amsthm}%
\usepackage{mathrsfs}%
\usepackage[title]{appendix}%
\usepackage{xcolor}%
\usepackage{textcomp}%
\usepackage{manyfoot}%
\usepackage{booktabs}%
\usepackage{algorithm}%
\usepackage{algorithmicx}%
\usepackage{algpseudocode}%
\usepackage{listings}%
\usepackage{lineno}
\usepackage{lscape}

\usepackage{hyperref}

\makeatletter
\gdef\orcidlogo{%
  \includegraphics[height=1.6ex]{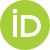}%
}
\makeatother

\theoremstyle{thmstyleone}%

\theoremstyle{thmstyletwo}%

\theoremstyle{thmstylethree}%

\begin{document}
\title[Transient narrowband radio bursts from 1E~1547.0$-$5408]{Transient narrowband radio bursts from the magnetar 1E~1547.0$-$5408}

\author*[1]{\fnm{Marcus~E.} \sur{Lower}\,\orcid{https://orcid.org/0000-0001-9208-0009}}\email{mlower@swin.edu.au}

\author[2]{\fnm{Paul} \sur{Scholz}\,\orcid{https://orcid.org/0000-0002-7374-7119}}

\author[3]{\fnm{Fernando} \sur{Camilo}\,\orcid{https://orcid.org/0000-0002-1873-3718}}

\author[4]{\fnm{David~M.} \sur{Palmer}\,\orcid{https://orcid.org/0000-0001-7128-0802}}

\author[5]{\fnm{John~E.} \sur{Reynolds}}

\author[6]{\fnm{John~M.} \sur{Sarkissian}}

\author[5]{\fnm{Lawrence~J.} \sur{Toomey}\,\orcid{https://orcid.org/0000-0003-3186-3266}}

\author[7,8]{\fnm{George} \sur{Younes}\,\orcid{https://orcid.org/0000-0002-7991-028X}}

\affil[1]{\orgdiv{Centre for Astrophysics and Supercomputing}, \orgname{Swinburne University of Technology}, \orgaddress{\street{PO Box 218}, \city{Hawthorn}, \postcode{3122}, \state{Victoria}, \country{Australia}}}

\affil[2]{\orgdiv{Department of Physics and Astronomy}, \orgname{York University}, \orgaddress{\street{4700 Keele Street}, \city{Toronto}, \postcode{MJ3 1P3}, \state{ON}, \country{Canada}}}

\affil[3]{\orgname{South African Radio Astronomy Observatory}, \orgaddress{\street{Liesbeek House, River Park}, \city{Cape Town}, \postcode{7700}, \country{South Africa}}}

\affil[4]{\orgname{Los Alamos National Laboratory}, \orgaddress{\street{PO Box 1663}, \city{Los Alamos}, \postcode{87545}, \state{NM}, \country{USA}}}

\affil[5]{\orgdiv{Australia Telescope National Facility}, \orgname{CSIRO, Space and Astronomy}, \orgaddress{\street{PO Box 76}, \city{Epping}, \postcode{1710}, \state{NSW}, \country{Australia}}}

\affil[6]{\orgdiv{Australia Telescope National Facility}, \orgname{CSIRO, Space and Astronomy}, \orgaddress{\street{Parkes Observatory, PO Box 276}, \city{Parkes}, \postcode{2870}, \state{NSW}, \country{Australia}}}

\affil[7]{\orgdiv{Astrophysics Science Division}, \orgname{NASA Goddard Space Flight Center}, \orgaddress{\street{8800 Greenbelt Road}, \city{Greenbelt}, \postcode{20771}, \state{Maryland}, \country{USA}}}

\affil[8]{\orgdiv{Center for Space Sciences and Technology}, \orgname{University of Maryland Baltimore County}, \orgaddress{\street{1000 Hilltop Cir}, \city{Baltimore County}, \postcode{21250}, \state{Maryland}, \country{USA}}}


\abstract{
Radio-loud magnetars are well known for exhibiting radiative behaviors that are seldom seen among the wider pulsar population. Yet one form of emission that remains elusive among pulsars and magnetars is narrowband bursts of radio waves. Such emission is a hallmark of repeating sources of fast radio bursts (FRBs), intense radio flashes that originate from distant galaxies. Here, we report the detection of 84 narrowband radio bursts during observations of the magnetar 1E~1547.0$-$5408 by the Murriyang telescope. They were confined to a transient profile component that appeared between 2009 February 23 to 25, one month after its 2009 outburst. Their appearance coincided with both dramatic changes in the magnetar line-of-sight magnetic-field geometry, and an emergent pulsed hard X-ray component detected by the Rossi X-ray Timing Explorer. The leading edge of the hard X-ray emission was phase-aligned with the narrowband component. This may indicate the bursts originated from pair cascades along closed field lines, though open-field line emission remains valid. Our characterization of the bursts suggests they may represent a low-energy analogue of the repeating FRB mechanism, further linking FRB progenitors to young, highly magnetized neutron stars.
}

\maketitle

1E~1547.0$-$5408 is a 2.07\,s spin-period pulsar that belongs to the rare class of radio-loud magnetars, neutron stars that are thought to be powered by the untwisting and decay of their ultra-strong magnetic fields \cite{Duncan1992, Paczynski1992}.
These objects have long been proposed to be the central engines of FRBs \cite{Bailes2022, Zhang2023}, an idea supported by the detection of an extraordinarily luminous radio burst from the Galactic magnetar SGR~1935$+$2154 in 2020 April \cite{CHIME2020b, Bochenek2020}.
1E~1547.0$-$5408 has been extensively monitored by Murriyang, the 64\,m Parkes radio telescope, following its 2007 discovery \cite{Camilo2008, Halpern2008, Lower2023a}.
It underwent a major outburst on 2008 October 3 \cite{Israel2010}, which was quickly followed by the largest outburst seen to date in this magnetar on 2009 January 22 \cite{Bernardini2011}.
Hundreds of hard X-ray/soft gamma-ray bursts were detected \cite{Mereghetti2009, Kaneko2010, Savchenko2010, vanderHorst2012}, along with a factor of $> 1000$ increase in soft X-ray flux \cite{Scholz2011} and an emergent hard X-ray component \cite{Enoto2010, Dib2012, Kuiper2012}.
As with the 2008 October and recent 2022 April outbursts \cite{Lower2023a}, the radio flux of 1E~1547.0$-$5408 was highly variable over the months after the 2009 outburst. 
Two bright, broadband radio bursts with a potential X-ray counterpart were detected during one of these periods of radio silence on 2009 February 3 \cite{Israel2021}.

These behaviors mirror that of SGR~1935$+$2154 throughout its 2020 April/October, and 2022 October outbursts \cite{Borghese22MNRAS, Younes2023, Hu2025}, thereby motivating searches for further FRB-like behavior from 1E~1547.0$-$5408.
Here, we focus on a subset of 13 radio observations of 1E~1547.0$-$5408 (four at 3094\,MHz and nine at 8356\,MHz, see Extended Data Table \ref{tab:obs}) taken with Murriyang between 2009 February 20 to March 1, along with X-ray observations taken by the Proportional Counter Array on board the Rossi X-ray Timing Explorer (RXTE; \cite{Jahoda1996}) spanning 2009 January 23 to July 6.
Details of the observations and data processing procedures can be found in the Methods.

Like other magnetars \cite{Kramer2007, Scholz2017, Lower2021a}, 1E~1547.0$-$5408 displayed clear variability in the number of detected radio profile components, flux density and polarization properties over time (Extended Data Figure \ref{fig:profiles}). 
During the 8356\,MHz observations taken between 2009 February 23 and 25, we detected intermittent pulses from a transient profile component that preceded the main profile by $\sim 70^{\circ}$--$100^{\circ}$ in pulse longitude.
In Figure 1 we present the time and frequency averaged polarization profile alongside the phase-resolved total intensity spectrum across the profile.
The standard component centered at phase $\sim 0^{\circ}$ displayed the characteristically flat and broadband spectrum that is common to magnetars. 
Whereas the offset transient component was comprised entirely of narrowband, short-duration bursts.

We visually identified a total of 84 narrowband bursts in the five 8356\,MHz radio observations covering this period (Extended Data Table \ref{tab:obs}), several examples of which are shown in Figure 2.
The alignment in pulse longitude, similar spectral modulations from diffractive scintillation to the persistent broadband emission, and dispersive sweep in the less narrow bursts indicate they are genuine astrophysical signals.
We also detected the bursts via a subband search of an independent data stream that recorded the individual pulses (Methods, see also Extended Data Figure~\ref{fig:afb}).
Only two pairs of bursts were found in the separate data stream to have occurred within the same rotation of the magnetar, indicating each detected burst was a distinct independent event.
The burst properties were characterized through a series of iterative temporal and spectral fits (Methods).
The resulting distributions of burst widths, spectral occupancies, center frequencies and fluences in Figure 3 do not appear to be strongly correlated with one another.
Most bursts were found to have recovered intrinsic widths that are below our $\sim$1-ms time sampling, indicating they are temporally unresolved.
We found no evidence of burst sub-structures that drift in frequency with time.

Our sample of narrowband bursts display a broad diversity in polarization fractions and position angle (PA) swings (Extended Data Figure \ref{fig:pol}).
The distribution of PA values appear to be clustered around a value of $\Psi \sim -28^{\circ}$, albeit with substantial scatter.
In contrast, the PA across the broadband emission initially followed an S-shaped sweep, reminiscent of the rotating vector model (RVM; Ref. \cite{Radhakrishnan1969}) for a simple dipole magnetic field geometry (Extended Data Figure \ref{fig:profiles}).
From February 23 onward, the PA sweep exhibited two discrete jumps, indicating the presence of orthogonally polarized modes (OPMs), which became gradually smoothed out over time.
The broader underlying changes in the PA direction and slope across the profile are consistent with changes in the magnetic geometry.
RVM fits to the data (Methods) show the magnetic inclination angle, PA inflection point and magnetic meridian display apparent cyclical variations with a period of $6.3 \pm 0.1$\,d (Extended Data Figure \ref{fig:rvm}).

We determined the absolute alignment of the radio and X-ray profiles shown in Figure 4 via phase-coherent timing using both the Murriyang and RXTE data (Methods).
Note, the alignment of the broadband component with the approximate mid-point of the soft (2--4\,keV) X-ray profile is consistent with that of a previous joint radio and X-ray observation by Murriyang and Chandra \cite{Israel2021}.
As indicated by the orange shading, the narrowband-emitting radio component overlaps in phase with the leading edge of the hard (10--33\,keV) X-ray profile.
This phase range also encompasses a narrow spike in the hard X-ray emission.
The aforementioned pair of broadband radio bursts detected at 6.6 GHz on 2009 February 3 appear to be similarly phase-aligned with the hard X-ray profile \cite{Israel2021}, and may have originated from the same emission region as the narrowband bursts.

Our discovery makes 1E~1547.0$-$5408 the fourth pulsar and second magnetar presently known to emit radio pulses with narrow spectral occupancies, with their appearance in the 8.1--8.6\,GHz band representing the highest frequency detection of narrowband emission from a neutron star to date.
A small fraction of giant pulses from the Crab pulsar and PSR~B0540$-$69 ($\sim 0.4\%$ and $\sim 2.8\%$ respectively) have similarly narrow bandwidths \cite{Thulasiram2021, Geyer2021}.
These `narrowband giants' are distinct from other spectral features that have been detected in giant pulses from several sources \cite{Bilous2015, Hankins2016, Mahajan2024}.
Transient narrowband emission was briefly detected from SGR~1935$+$2154 following a possible anti-glitch and intermediate-luminosity radio burst in 2020 October \cite{Younes2023, Giri2023, Zhu2023, Wang2024}.
Unlike decade-long stability of giant pulse emission from the Crab and PSR~B0540$-$69, the narrowband profile components of 1E~1547.0$-$5408 and SGR~1935$+$2154 were extremely short-lived, appearing for only 3\,d and 22--27\,d respectively.
Both emerged soon after both spin-down dominated timing events and a decline in their high-energy burst activity \cite{Dib2012, Younes2023}.
In the case of 1E~1547.0$-$5408, the X-ray burst counts from the Fermi Gamma-ray Burst Monitor and Swift Burst Alert Telescope indicate the high-energy burst rate had dropped precipitously by 2009 February 24 \cite{vonkienllin20ApJ:gbm}.
Similarly, the FRB-like bursts from SGR~1935$+$2154 followed drops in activity during the 2020 April and 2022 October outbursts \cite{younes20ApJ:1935, Borghese22MNRAS, Younes2023, Hu2024}.

Radio emission from magnetars is thought to originate from charged particles that are either accelerated within bundles of closed field lines \cite{Beloborodov2009} or open field lines that become twisted during an outburst \cite{Szary2015, Zeng2026}.
The detected radio pulses are typically broadband and persist for several years, unlike the short-lived narrowband components seen in both 1E~1547.0$-$5408 and SGR~1935$+$2154.
As shown in Figure~\ref{fig:xray}, the broadband component of 1E~1547.0$-$5408 leads the soft (2--4\,keV) X-ray peak by $\sim100^{\circ}$ in pulse longitude. 
Similar offsets ranging between $36^{\circ}$--$144^{\circ}$ have been observed among other radio-loud magnetars \cite{Pennucci2015, Gotthelf2019, Bansal2023}.
Radio emission from either closed or high-altitude field lines can lead to such offsets, as can surface hotspots that are misaligned with the active magnetic pole \cite{Gotthelf2019, Stewart2025}.
The narrowband component was phase aligned with the leading edge of its hard (10--33\,keV) X-ray profile, both of which were near-anti-aligned with the soft X-ray peak.
X-rays detected from magnetars at energies up to 150--200\,keV are believed to originate from resonant inverse Compton scattering by charged particles traveling along closed magnetic field lines \cite{Baring2007, Fernandez2007, Beloborodov2013a, Wadiasingh2018}, or via synchrotron emission from electron/positron cascades along closed field loops \cite{Baring2001, Harding2025}.
While there is a non-negligible $\sim 11$--$38\%$ probability that these two emission regions overlap one another by chance (Methods), it is worth noting that SGR~1935$+$2154 displayed a marginally detected 5--10\,keV pulsed X-ray component at the same phase range as its narrowband radio pulses, both of which were similarly anti-aligned with the soft X-ray peak \cite{Borghese22MNRAS, Zhu2023}.
While a chance coincidence in 1E~1547.0$-$5408 cannot be definitively ruled out, observations of phase-aligned narrowband radio and hard X-ray components in two separate magnetars may point to a shared emission region \cite{Harding2025}.
The implied production of coherent radio bursts from closed magnetic field lines could help explain the lack of apparent millisecond to second-scale rotation periods associated with repeating FRB sources (e.g. Ref. \cite{DiLi2021}), as the emission is not necessarily beamed along the narrow open field lines of the dipole axis.
However, we note narrowband emission originating from the open field lines above the polar cap cannot be ruled out. 
Given the near-aligned magnetic and viewing geometry of 1E~1547.0$-$5408 \cite{Stewart2025}, the angular offset of the narrowband emission component from the magnetic axis is only $\sim 16^{\circ}$ (Methods).
Transient emission from an outer part of the open-field region, combined with differences in magnetic and viewing geometry, could explain the lack of detected broadband radio pulses from SGR~1935$+$2154.

The similarity in emission bandwidth to bursts from repeating FRB sources could be indicative of a shared emission mechanism that operates over a broad range of energies.
For instance, the weakest burst we detected from 1E~1547.0$-$5408 had a fluence of 1.3 Jy ms, while the brightest was 412.3 Jy ms, which at the nominal distance of 4.5 kpc \cite{Tiengo2010} correspond to pseudo luminosity range of $L_{\nu} = 5.0 \times 10^{20} $--$1.6 \times 10^{23}$\,erg\,s$^{-1}$\,Hz$^{-1}$.
This is an order of magnitude weaker than the broadband radio bursts detected from 1E~1547.0$-$5408 on 2009 February 3 \cite{Israel2021}, and about five to eight orders of magnitude less luminous than the nearest extragalactic repeating FRB \cite{Nimmo2023}.
Their energetics are comparable to the weakest of the sporadic broadband bursts from SGR~1935$+$2154 \cite{Kirsten2021, Giri2023}, and approximately 1--3 orders of magnitude higher than its narrowband pulses \cite{Zhu2023, Wang2024}. 

Despite the difference in energy, the distributions of burst durations and bandwidths in Figure 3 are comparable to those of several repeating FRB sources.
This includes rapid changes in PA swing (Extended Data Figure \ref{fig:rvm}) that have been interpreted as evidence in favor of a magnetospheric origin for FRBs \cite{Luo2020}.
Both FRBs 20121102A and 20240114A exhibit a narrowing of burst widths at redshift-corrected rest-frame frequencies of 4.7--9.5\,GHz and 1.6--7.6\,GHz respectively \cite{Gajjar2018, Limaye2026}.
Several repeating FRBs also display burst sub-structures with intrinsic widths in the microsecond to nanosecond regime \cite{Nimmo2021, Hewitt2023, Snelders2023}.
Hence, the sub-ms burst durations among our sample may not be surprising.
Similarly, the drift rates of repeating FRBs has been found to follow a positive power-law relation with observing frequency \cite{Chamma2023}.
At 8356\,MHz, the expected drift rate would be between 100's--1000's of MHz\,ms$^{-1}$, which is imperceptible given our 1\,ms time sampling and explains the lack of detected drifting sub-structures
Unresolved drifting sub-structures could explain the multi-peaked spectra seen among some bursts.
This link between repeating FRBs and magnetars is strengthened by a fraction of the narrowband pulses from SGR 1935+2154 displaying downward frequency drifts \cite{Zhu2023}, with reported drift rates of $\sim 10$\,MHz\,ms$^{-1}$ at 1.5 GHz \cite{Wang2024}.
This is consistent with the drift rates of repeating FRBs observed at similar frequencies. 
The high burst rates and short detection windows of the bursts from 1E~1547.0$-$5408 and SGR~1935$+$2154 are reminiscent of burst storms among FRBs 20121102A and 20200120E \cite{Hewitt2022, Nimmo2023}, and periods of hyperactivity seen in several repeaters that last several days to months \cite{Konijn2024, Bilous2025, Shin2026, Shaji2026}.
Similarly sudden quenching of emission within a single day has also been observed in several active periods of FRBs 20201124A and 20240619D \cite{Xu2022, Zhou2022, Shaji2026}.

The narrowband spikes and double-peaked spectra observed among the bursts in Figure 2 appear qualitatively similar to the sharp spectral features predicted to result from caustics in plasma lenses \cite{Cordes2017}.
However, this model faces several challenges in the context of 1E~1547.0$-$5408 (see Methods for details).
A putative lens would have to be relatively compact and confined to the co-rotating magnetosphere in order to not affect the broadband profile component. 
The relativistic nature of the magnetospheric plasma would suppress dispersive effects, meaning a high plasma density is required to produce caustics in he 8--9\,GHz band.
Combined with a finely-tuned lens shape/viewing geometry, a plasma lensing origin for the narrowband structures appears disfavored.
Synchrotron absorption is also unable to fully explain the observed narrowband and band-limited features given the high radio frequency of the bursts \cite{Wang2022}. 
Rather, the narrowband nature of these bursts is likely a feature of their emission mechanism.

The narrowband profile component of 1E~1547.0$-$5408 emerged during a period of extreme polarimetric variability across the more persistent broadband emission.
Similar such variations have been reported in the radio magnetars XTE~J1810$-$197 and Swift J1818.0$-$1607, which were respectively interpreted as originating from damped free precession or shifts in the radio emission region \cite{Desvignes2024, Lower2021a}.
The apparent cyclical changes in $\alpha$ and $\phi_{0}$ (Extended Data Figure \ref{fig:rvm}) are consistent with free-precession induced changes in the longitude and latitude of the magnetic axis \cite{Gao2023}, where the brief appearance of the narrowband profile component could be a result of the corresponding active region crossing our line of sight as the magnetar precesses.
The precession in XTE~J1810$-$197 was suggested to have been triggered by a deformation of the crust associated with the 2018 November outburst, that quickly relaxed back to the pre-outburst axisymmetry \cite{Desvignes2024}.
For a potential precession period of $\sim 6.3$\,d, we infer a neutron star ellipticity of $\epsilon = P_{\rm spin}/P_{\rm precession} \sim 3.8 \times 10^{-6}$.
This is close to the maximum inferred value for XTE~J1810$-$197 and consistent with models of crustal elasticity and magnetic deformation in neutron stars \cite{Johnson-McDaniel2013, Haskell2008}.
Free precession had previously been invoked as an explanation for both the 16.4\,d periodicity in the activity of FRB~20180916B and the shifts in emission frequency observed within each activity window \cite{Levin2020, Li2021}, along with the the longer $\sim 157$\,d activity cycle in FRB~20121102A \cite{Rajwade2020}.
However, more recent analyses of the PA variability of FRB~20180916B disfavor free-precession-based models where a neutron star with a $\sim$hour-duration rotation period precesses at a rate that matches the periodicity of the activity cycle \cite{Bethapudi2025}.
Additionally, internal dissipative processes are expected to damp free-precession on relatively short timescales in magnetars \cite{Lander2018}.
Such behavior was reported in XTE~J1810$-$197 \cite{Desvignes2024}, and likely also took place in 1E~1547.0$-$5408 on a similarly short $\sim$month-long timescale.
This would be consistent with the reported long-term stability in its magnetic and viewing geometry \cite{Lower2023a, Stewart2025}.
Phase coherent timing of the 1E~1547.0$-$5408 may offer a means to differentiate between these interpretations, as free precession is expected to introduce periodic changes in the magnetar spin-down rate on the same timescale as the polarimetric variations.

Alternatively, the polarimetric variations could instead result from plastic motion of the neutron star crust as internal energy dissipates following the outburst \cite{Jones2003, Lyutikov2015}.
This mechanism has been suggested as a means of producing the longitudinal drifts in the X-ray emitting hotspots on the surfaces of SGR~1830$-$0645 and 1E~1841$-$045 \cite{Younes2022, Younes2025}.
Changes in the magnetic field structure due to shifts in the magnetic footprints would naturally alter the magnetic topology within the radio emission region and the observed polarization state. 
However, the linear shifts in the X-ray profile features of SGR~1830$-$0645 and 1E~1841$-$045 occurred over several months, as opposed to day-timescale oscillations we observed in 1E~1547.0$-$5408.
Propagation through a magnetized plasma along the line of sight has also been proposed to produce RVM-like PA sweeps independent of magnetic geometry \cite{Benacek2025}, and was previously invoked to explain both the complex PA sweep and coincident conversion of linear to circular polarization in the radio emission of XTE~J1810$-$197 \cite{Lower2024}.
In the case of 1E~1547.0$-$5408, the 6.3\,d period may be due to oscillations in the plasma density along the line of sight.
Crustal displacement or plasma propagation, if confirmed, would provide novel insights into the dynamics of magnetars and how their magnetic fields couple to the emitted radiation.
Such behaviors would also need to be considered when attempting to understand the polarization evolution of FRB sources over short and long timescales.

The discovery that several nearby neutron stars produce narrowband radio emission akin to repeating FRBs raises the tantalizing prospect of a shared emission mechanism, albeit one that must operate over a vast range of energy scales.
Future observations of this phenomena with the latest generation of wideband receivers at sub-ms timescales will reveal if they follow the same burst duration, bandwidth and drift-rate relations seen among repeating FRB sources.
Detections from other magnetars, both radio loud and seemingly radio-silent, alongside coincident observations by hard X-ray telescopes are crucial to resolving role of pair cascades in closed or open magnetic field lines in their production.

\bibliography{sn-bibliography}

\clearpage

\begin{figure}
    \centering
    \includegraphics[width=0.9\linewidth]{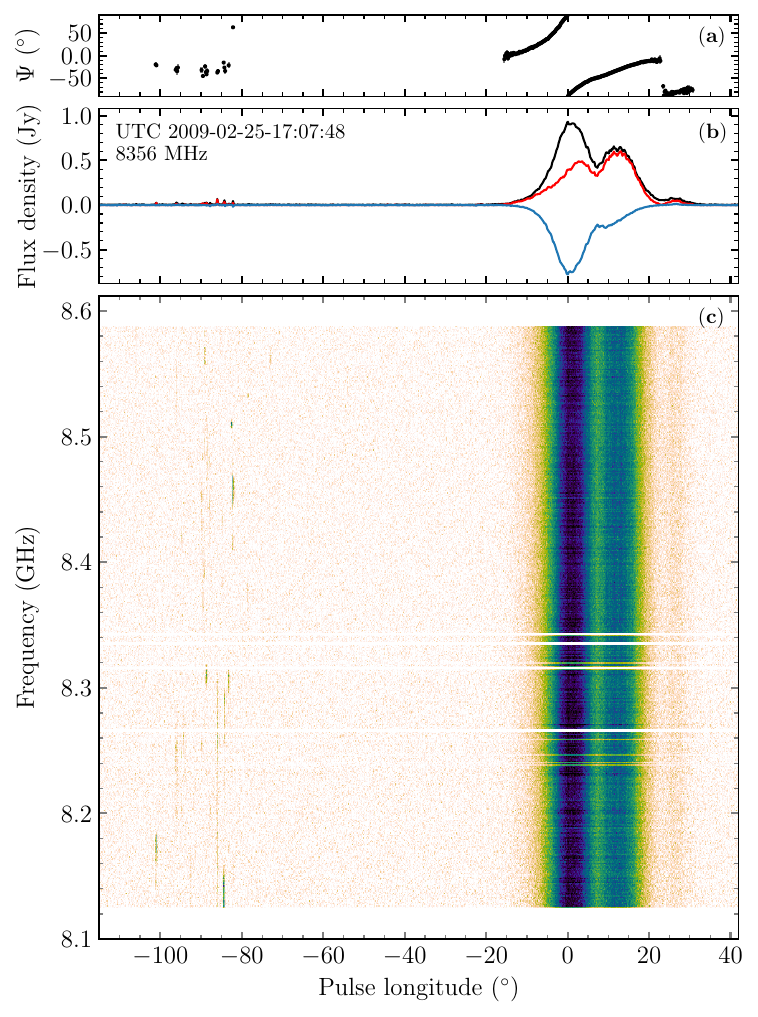}
    \caption{{\bf Detection of narrowband and broadband radio emission from 1E~1547.0$-$5408 at 8356\,MHz}. Panels show the linear polarization position angle swing ($\Psi$) across the profile ({\bf a}), polarization profile with total intensity in black, linear polarization in red and circular polarization in blue ({\bf b}), and phase-resolved total intensity spectrum ({\bf c}). The time-stamp in panel ({\bf b}) indicates the UTC start time of the observation. Horizontal gaps in panel ({\bf c}) indicate channels excised due to radio interference. Note the dynamic range of the bottom panel has been reduced to highlight low-intensity features.}
    \label{fig:waterfall}
\end{figure}

\clearpage

\begin{figure*}
    \centering
    \includegraphics[height=0.316\linewidth]{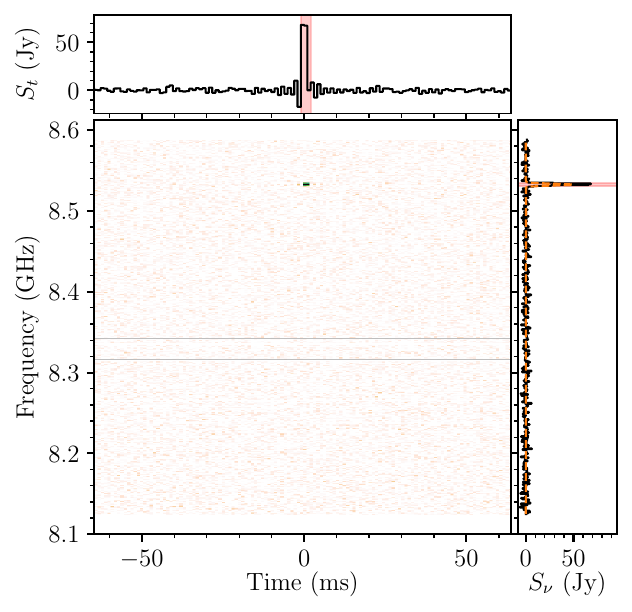}
    \includegraphics[height=0.316\linewidth]{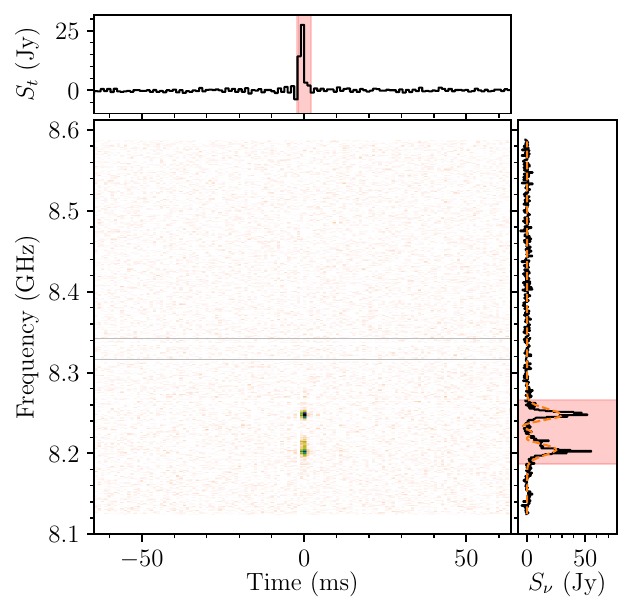}
    \includegraphics[height=0.316\linewidth]{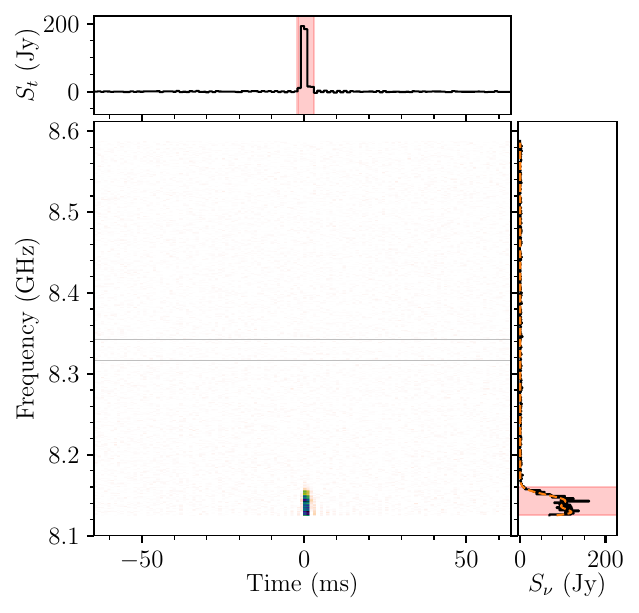}
    \includegraphics[height=0.316\linewidth]{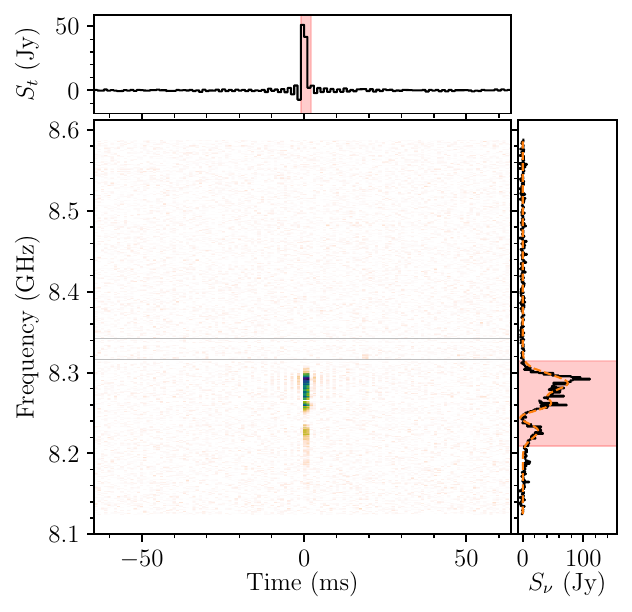}
    \includegraphics[height=0.316\linewidth]{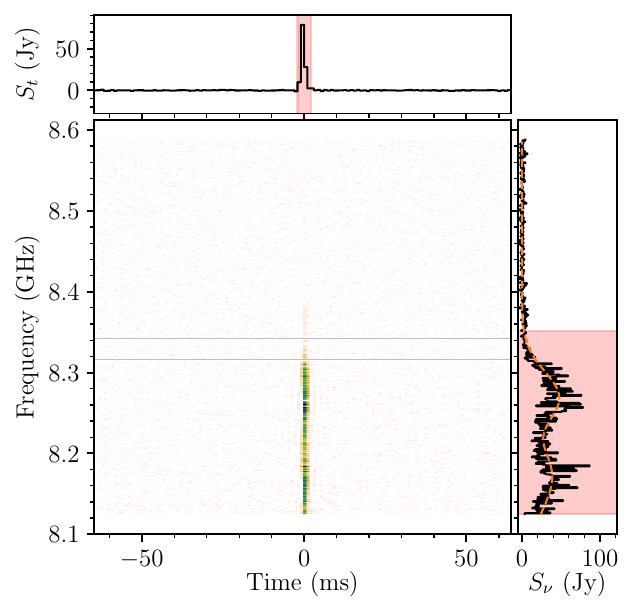}
    \includegraphics[height=0.316\linewidth]{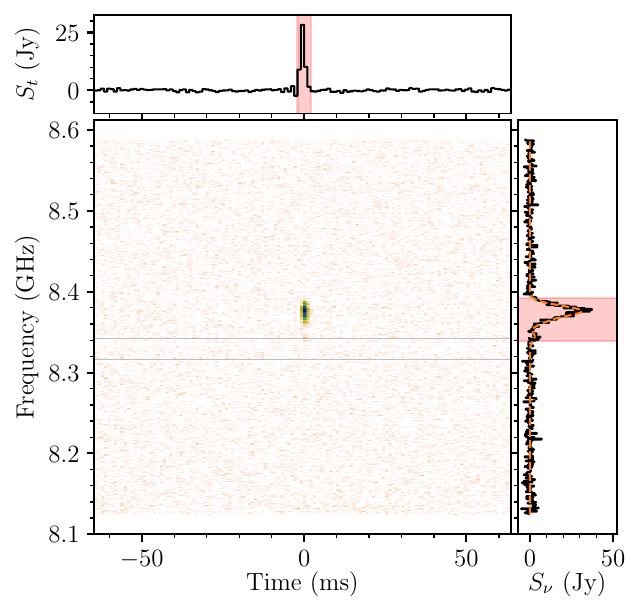}
    \includegraphics[height=0.316\linewidth]{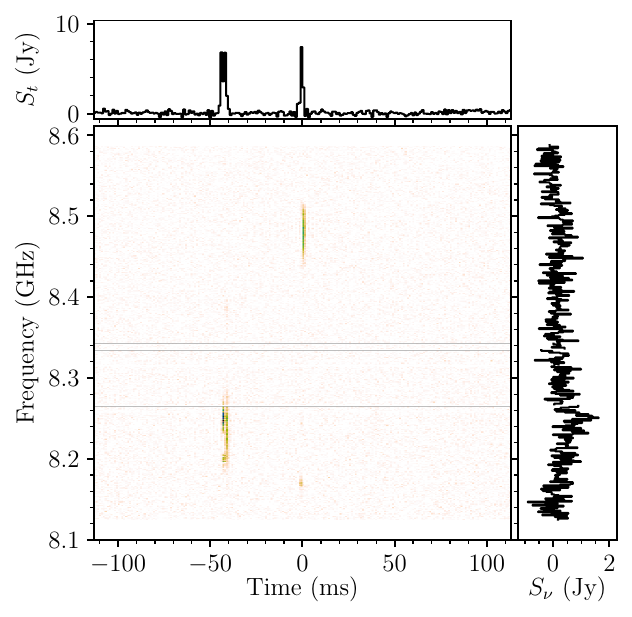}
    \includegraphics[height=0.316\linewidth]{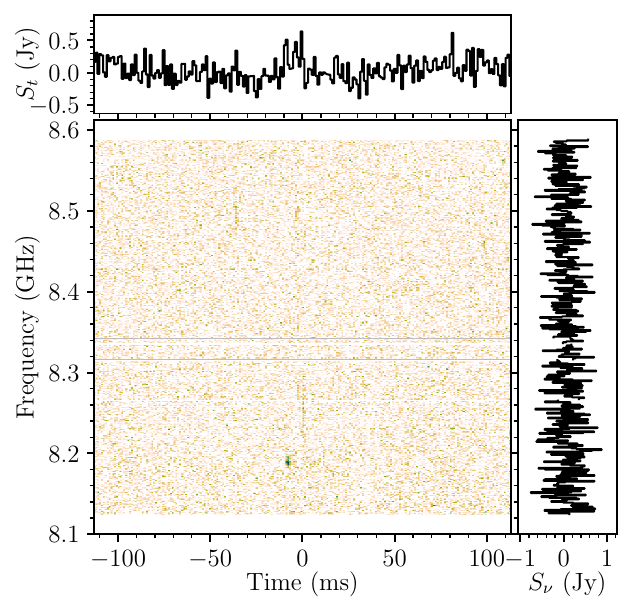}
    \includegraphics[height=0.316\linewidth]{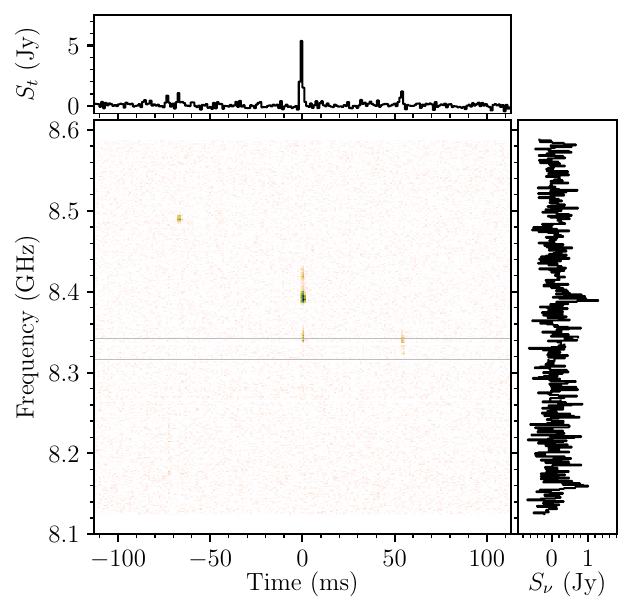}
    \caption{{\bf Examples of narrowband bursts detected from 1E~1547.0$-$5408.} Sub-integrations containing only a single narrowband burst (top and middle) and those containing multiple bursts (bottom). The middle panels of each plot shows the burst dynamic spectrum. Upper panels the frequency-averaged profile of the bursts, while the red shaded region is the time-interval encompassing the burst. Right-hand panels show the same as the upper panels but for the burst spectra, where the orange traces indicate the median recovered fit to the spectrum.}
    \label{fig:bursts}
\end{figure*}

\clearpage

\begin{figure}
    \centering
    \includegraphics[width=0.9\linewidth]{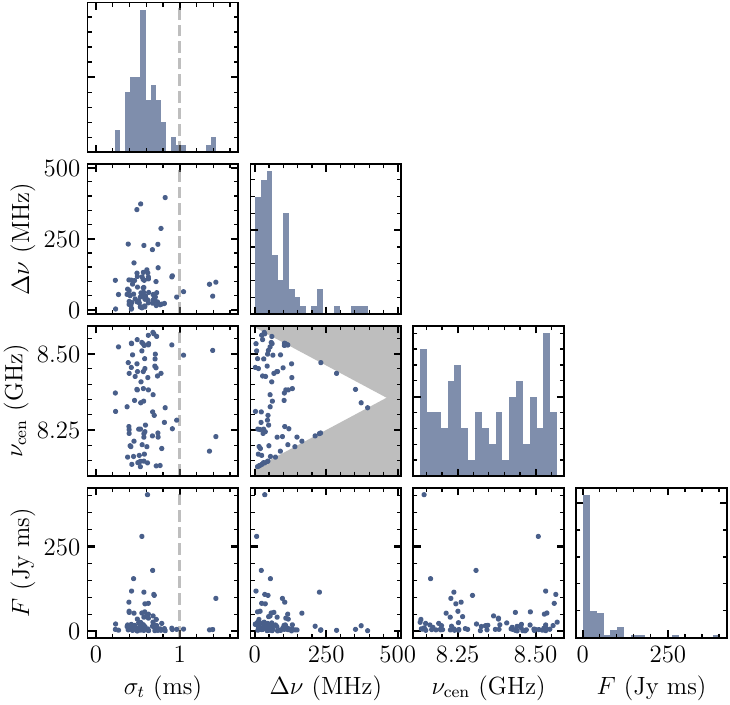}
    \caption{{\bf One and two-dimensional distributions of the narrowband burst properties}. Panels contain the recovered burst widths ($\sigma_{t}$), spectral occupancy ($\Delta\nu$), central frequency ($\nu_{\rm cen}$) and fluence ($F$). The vertical dashed gray line indicates the $\sim$1\,ms time resolution of the data. Shaded gray region indicates the observing band limits on the inferred spectral occupancy and center frequency.}
    \label{fig:corner}
\end{figure}

\clearpage

\begin{figure}
    \centering
    \includegraphics[width=0.9\linewidth]{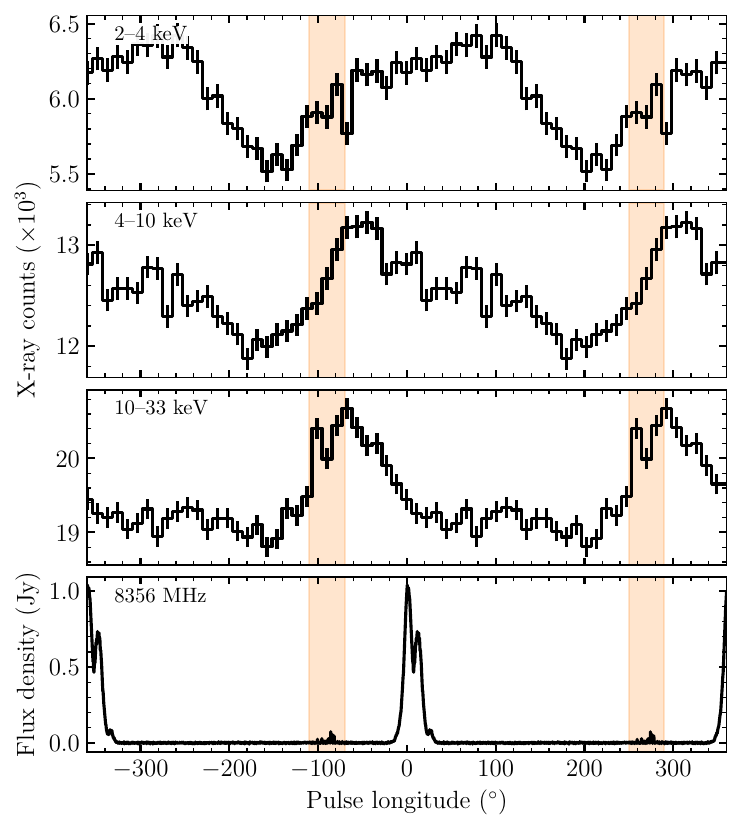}
    \caption{{\bf Comparison of the X-ray and radio profile of 1E~1547.0$-$5408}. The X-ray data are split into three different energy bands while the total intensity radio profile is the same as that shown in Figure~\ref{fig:waterfall}. The orange shaded region indicates the pulse longitude range over which all 84 narrowband bursts were detected. X-ray error bars were derived assuming Poisson uncertainties on the per-bin counts. Two rotations of the magnetar are displayed for clarity.}
    \label{fig:xray}
\end{figure}

\clearpage

\section{Methods}\label{sec11}

\subsection{Radio and X-ray observations}

There were 13 observations of 1E~1547.0$-$5408 that were collected between 2009 February 20 to March 1 (MJD 54882--54891) by the 64\,m Murriyang telescope at the Parkes Observatory.
The first four observations were conducted at 3094\,MHz using the 10/50-cm receiver \cite{Granet2005}, and the remaining nine were taken at 8356\,MHz with the MARS X-band receiver.
These observations vary in length from 58\,s to 1819\,s and were conducted as part of a wider monitoring campaign triggered by the 2009 January 22 outburst \cite{Bernardini2011}.
A complete characterization of the long-term rotational and emission behavior of 1E~1547.0$-$5408 will be presented in a future work (Camilo et al. in prep.).
Pulsar fold-mode data with 20--30\,s sub-integrations containing 1024--2048 phase bins were recorded using the Pulsar Digital Filterbank v4 (DFB4) signal processor, with 1\,MHz wide frequency channels covering 1024\,MHz of bandwidth for the 10/50cm receiver and 512\,MHz for MARS, and saved to {\tt psrfits} format archive files \cite{Hotan2004}.
The center frequency, duration and sub-integration lengths of the individual fold-mode observations are recorded in Extended Data Table~\ref{tab:obs}.

The radio data were flux and polarization calibrated using off-source observations of a pulsed noise diode and on/off observations of Hydra A as a flux density reference using the {\tt pac} tool in {\tt psrchive} \cite{Hotan2004, vanStraten2012}.
We verified the polarization calibration of the MARS data by applying the same reduction procedure to two near-in-time observations of the bright radio pulsar PSR~B1641$-$45 that were taken on 2009 February 24 and 25. 
A comparison between our recovered polarization profiles for PSR~B1641$-$45 and the reference observation from Ref. \cite{Johnston2006} can be found in the Supplementary Materials.
The polarization profiles we recover are in good agreement with the reference observation.
The profiles were aligned based on an ephemeris derived from X-ray timing of 1E~1547.0$-$5408 (see below).
Simultaneous pulsar search-mode data with the MARS receiver were also recorded using an analogue filterbank (AFB; Ref. \cite{Manchester2001}) system with 1-bit digitization, 1\,ms sampling and 192 channels each 3\,MHz wide.
We folded the AFB search data at the rotation period of the magnetar using {\tt dspsr} \cite{vanStraten2011} and saved the resulting single-pulse sub-integrations to {\tt psrfits} format archives.
No attempt was made to calibrate the AFB data.
Frequency channels contaminated by radio-frequency interference (RFI) were excised from both the AFB and DFB4 data manually using the {\tt pazi} tool from {\tt psrchive}.
Both data sets were dedispersed using a dispersion measure of 697\,pc\,cm$^{-3}$. 
The DFB4 linear polarization data were de-Faraday rotated to infinite frequency by a rotation measure of $-1847.6$\,rad\,m$^{-2}$ \cite{Lower2023a} using the {\tt --aux\_rm} flag in the {\tt pam} tool of {\tt psrchive}.

We also use X-ray observations obtained using the Proportional Counter Array (PCA) instrument on the Rossi X-ray Timing Explorer (RXTE; Ref. \cite{Jahoda1996}) spanning 2009 January 23 to 2009 July 6 (MJD 54854--55018). 
PCA observations were collected in \textit{GoodXenon} mode which records photon arrival times to 1\,$\mu$s precision and bins photon energies into 256 channels over PCA's 2--60\,keV band. 
Photon arrival times were corrected to the solar system barycenter using a source position of RA, Dec (J2000) = 15:50:54.11, $-$54:18:23.7 \cite{Gelfand2007}. 
As 1E~1547.0$-$5408 was emitting a high rate of short X-ray bursts following its 2009 January outburst, we first searched for these bursts to remove them from our timing and pulse profile analysis. We searched time-series in two energy bands, 2--10 and 10--33\,keV, binned at 0.03125\,s time resolution. 
We compared time-series by comparing each bin to a running mean calculated over a 16\,s window centered on the search bin, as the background count rate varies significantly over the observations. 
If the Poisson probability of the number of counts in a bin exceeded $0.01/N$ (where $N$ is the total number of bins searched) given the measured local mean count rate, it was flagged as a burst and photons from that time span were removed from our subsequent analyses.

\subsection{Alignment of the radio and X-ray profiles}

For our timing analysis, we used RXTE photons from the 2--10\,keV band. A template profile was constructed by folding the 2--10\,keV photons from all observations from 2009 Jan 22 to 2009 Dec 26 \cite[listed in Table 2 of][]{Dib2012} by the post-2009 outburst ephemeris in Ref. \cite{Dib2012}. Pulse times-of-arrival (ToAs) for RXTE observations were extracted using the maximum likelihood method described in Ref. \cite{Scholz2012} using two harmonics. Four ToAs were extracted per observation by deriving a ToA from each quarter of the photons per observation. 

Using {\sc tempo2} with pulse-numbering enabled \cite{Hobbs2006}, we obtained a phase-coherent timing solution covering RXTE ToAs between 2009 January 23 to July 6 by fitting for the magnetar spin-frequency along with its first and second derivatives.
We then whitened the residuals by fitting a series of harmonically related sinusoids to the ToAs using the {\sc fitwaves} plugin to {\sc tempo2} \cite{Hobbs2004}.
ToAs were derived from the radio observations by cross-correlating time and frequency averaged copies of the total intensity profiles with a noise-free Gaussian template.
We then re-fit the whitened timing solution to the radio ToAs to determine the zero-phase point of the magnetar at the reference epoch UTC 2009-02-23-23:54:45.678 (the start of the X-ray timing observations).
After flagging data containing short X-ray bursts, we then folded the RXTE photons collected in observations taken between 2009 Feb 6 to March 16 using the whitened ephemeris and the {\tt photonphase} tool in {\tt PINT} \cite{Luo2021}.
We show the resulting phase-aligned X-ray profile split into 2--4\,keV, 4--10\,keV and 10--33\,keV bands in Figure~\ref{fig:xray} alongside the UTC~2009-02-25:17:07:48 radio observation. 
The overall X-ray profile shapes are broadly consistent with those reported by over a similar timespan and energy range \cite{Dib2012, Kuiper2012}.

We inferred the chance probability that the phase range occupied by the narrowband radio component overlapped with the hard X-ray profile component as
\begin{equation}
    P({\rm Overlap}) = (W_{\rm NB} + W_{\rm X})/360.
\end{equation}
Here, $W_{\rm NB} \sim 40^{\circ}$ is the width of the narrowband radio component and $W_{\rm X} \sim 100^{\circ}$ the width of the hard X-ray component, from which we obtain a chance overlap probability of $P({\rm Overlap}) \approx 0.38$ or 38\%.
We also computed the chance probability that a radio component would fall at the observed offset from the hard X-ray peak as
\begin{equation}
    P(|\delta\phi_{\rm rand}| \leq |\delta\phi_{\rm rand}|) = \int_{-|\delta\phi_{\rm obs}|}^{+|\delta\phi_{\rm rand}|} f(\phi) d\phi,
\end{equation}
where $|\delta\phi_{\rm rand}|$ is the offset of a randomly generated radio component and $|\delta\phi_{\rm obs}| = 0.056$ is the observed phase offset.
Assuming a radio component has a uniform probability of appearing at any rotation phase, this equation simplifies to $P(|\delta\phi_{\rm rand}| \leq |\delta\phi_{\rm rand}|) = 2|\delta\phi_{\rm obs}| \approx 0.11$ or 11\%.

\subsection{Identification of the narrowband bursts}

The narrowband bursts recorded by the DFB4 signal processor were readily identifiable through visual inspection of dynamic spectra produced from the individual 20--30\,s duration sub-integrations recorded at each epoch.
From this alone we identified 84 bursts.
To verify the existence of these events, we performed a subband search for bursts in the AFB data.
The individual archives were each split into two, four and eight subbands, with respective bandwidths of 96, 48 and 24\,MHz.
After averaging each subband in frequency, we performed a matched-filter search by cross-correlating the resulting timeseries with a 2\,ms wide boxcar (equivalent to two time samples in width).
Wider boxcars did not produce any clear, positive candidates upon inspection and reduced the signal-to-noise ratio (S/N) of visually identifiable bursts.
We then inspected all candidates with a S/N $\geq$ 5, where out of 72 burst candidates, 69 were identified as likely being real bursts.
The remaining three candidates appeared indistinguishable from the surrounding noise.
Note, that the lower number of bursts identified among the AFB data is likely due to both the weaker and most band-limited bursts being indistinguishable from noise in the AFB.

\subsection{Burst characterization}

We characterized the properties of the 84 bursts found in the DFB4 data stream by conducting a series of iterative fits.
In order to extract frequency-averaged profiles or time-averaged spectra, we placed `on-pulse' windows around the bursts in their dynamic spectra.
The upper and lower bounds of these windows were initially determined by eye, and subsequently refined following the procedures described below.
For our temporal fits, we found that the majority of the burst profiles were well described by a single Gaussian of the form
\begin{equation}
    S(t) = \frac{A}{\sqrt{2\pi\sigma_{t}^{2}}} \exp \Big[ -\frac{(t - \mu_{t})^{2}}{2\sigma_{t}^{2}} \Big],
\end{equation}
where $A$, $\sigma_{t}$ and $\mu_{t}$ are the amplitude, width and centre (mean) of the Gaussian.
Two distinct Gaussian sub-components were needed in one case to fully capture the burst profile.
We used the Bayesian Inference Library, {\tt bilby} \cite{Ashton2019}, to sample the model parameters with the {\tt dynesty} nested sampling algorithm \cite{Speagle2020}, assuming uniform priors and a Gaussian likelihood function of the form
\begin{equation}\label{eqn:like}
    \mathcal{L}(d | \vec{\theta}) = \prod_{i}^{n} \frac{1}{\sqrt{2\pi\sigma^{2}}} \exp \Big[ -\frac{(d_{i} - \mu(\vec{\theta}))^{2}}{2\sigma^{2}} \Big].
\end{equation}
Here, $d$ is the data being fit, $\vec{\theta}$ is a vector containing the model parameters, $\sigma$ is a free parameter used to constrain the scatter in the data, and $\mu(\vec{\theta})$ is the model that is being fit to the data.
We then determined the burst durations by computing the 5\,percent width of the median a-posteriori Gaussian fit to the data. 
This on-pulse interval was extended by $\pm 2$\,ms when extracting the burst spectra.

We fit the spectral response of the bursts using a shapelet-based spectral model \cite{Refregier2003}
\begin{equation}
    S_{\nu} = A \sum_{n = 0}^{N_{s}} C_{n}\, \omega^{-1/2} \phi_{n}\Big( \frac{\nu - \nu_{0}}{\omega} \Big),
\end{equation}
where $A$ is the normalizing amplitude, $N_{s}$ is the number of shapelet components, $C_{n}$ is the shapelet component amplitude, $\omega$ the width and $\phi_{n}$ is the basis function given by
\begin{equation}
    \phi_{n}(\nu) = \frac{H_{n}(\nu) e^{-\nu^{2}/2}}{\sqrt{2^{n} \sqrt{\pi}\, n!}},
\end{equation}
in which $H_{n}(\nu)$ is a Hermite polynomial of $n$-th degree.
Five shapelet components were generally sufficient to capture intra-burst spectral variations, except for the narrowest bursts for which only a single component shapelet (approximating a Gaussian) was required.
In several instances the burst spectra were composed of multiple components separated by 10's to 100's of MHz, necessitating the addition of two independent shapelet profiles. 
As with the temporal fits, we used the same Gaussian likelihood function in Equation~\ref{eqn:like} when sampling the shapelet parameters.
We then derived the spectral occupancies from the radio frequency range encompassed by the 95\% upper and lower bounds of the median a-posteriori spectral model.
Burst fluences were computed by summing over the on-pulse regions of each burst, multiplied by the sampling time ($\Delta t$) as 
\begin{equation}
    F = \sum_{i}^{N_{\rm on}} S_{t,i} \Delta t,
\end{equation}
where $N_{\rm on}$ is the number of on-pulse bins.

\subsection{Polarization properties and geometry}

Under the assumption that the radio emission originates from above the polar cap of a dipole magnetic-field, the PA swing can be modelled via the RVM as
\begin{equation}
    \tan(\Psi - \Psi_{0}) = \frac{\sin\alpha \sin(\phi - \phi_{0})}{\sin\zeta \cos\alpha - \cos\zeta \sin\alpha \cos(\phi - \phi_{0})},
\end{equation}
where $\alpha$ is the inclination angle between the magnetic and spin axes, $\zeta$ is the viewing angle from the spin axis, and $\Psi_{0}$ is the inflection point of the PA sweep which occurs at a reference phase of $\phi_{0}$. 
The angle between the magnetic axis and our line of sight can be derived as $\beta = \zeta - \alpha$.
Note, that in our observer reference frame that the PA will be inverted, i.e $\Psi_{\rm obs} = -\Psi$.
In order to avoid possible model misspecification when modeling the 8356\,MHz PA swings, we ignored the PA within $\pm5^{\circ}$ around the first OPM.
Using the same Bayesian inference framework applied to the spectral modeling, we sampled the RVM parameters by fitting the PAs derived from each epoch.
Uniform priors were assumed for $\Psi_{0}$ and $\phi_{0}$. 
We used sine priors for $\alpha$ and $\zeta$ to ensure that these parameters were sampled uniformly in solid angle.
The RVM fits were generally well-matched to the data, with median a-posteriori reduced chi-square values ranging between $\chi_{\nu}^{2} = 0.71$ to $1.61$. 
However, the recovered values of $\alpha$, $\zeta$, $\Psi_{0}$ and $\phi_{0}$ varied by up to tens of degrees from one epoch to the next, following an overall sinusoidal trend.
Large cyclical changes in $\zeta$ over time are difficult to physically reconcile, as it would imply the rotation axis of the magnetar is precessing about a misaligned total angular momentum vector.
While this can arise from spin-orbit coupling in a relativistic binary system, it would persist over many years. 
1E~1547.0$-$5408 also shows no evidence of having a binary companion.
Additionally, values of $\zeta$ that deviate from near-alignment with the rotation axis are also inconsistent polarimetric studies of 1E~1547.0$-$5408 during periods of relative stability, where it was found to be consistent with an aligned rotator viewed close to the rotation axis \cite{Camilo2008, Stewart2025}.
The 2-D posterior distributions for $\alpha$ and $\zeta$ were strongly correlated at all epochs.
Hence, physical changes in $\alpha$ alone could be manifesting as an apparent variability in $\zeta$.
We therefore conducted an additional set of RVM fits to the data using a restricted Gaussian prior of $\pi(\zeta) = \mathcal{N}(7.5^{\circ},2.9^{\circ})$ to approximate the posterior distribution on $\zeta$ recovered by Ref. \cite{Stewart2025} for observations taken in 2025 March.
Near equivalent values of $\chi_{\nu}^{2}$ were obtained at each epoch to the unrestricted RVM priors, indicating the restricted $\zeta$ prior returned similarly statistically acceptable fits to the data.

We can compute the angular offset between the magnetic axis of the magnetar and our line of sight throughout a full rotation via simple spherical geometry.
The vectors encoding the position of the magnetic moment and our line of sight on unit circle are given by $\vec{\mu} = \{\cos\phi(t)\cos\alpha, \sin\phi(t)\cos\alpha, \sin\alpha\}$ and $\vec{\ell} = \{\cos\phi_{0}\cos\zeta, \sin\phi_{0}\cos\zeta, \sin\zeta\}$.
We can then infer the angle separating these two vectors as $\cos\theta_{\mu\rightarrow\ell} = (\vec{\mu} \cdot \vec{\ell})/(|\vec{\mu}| \times |\vec{\ell}|)$. 
Taking the inferred magnetic and viewing geometry recovered by our RVM fit to the UTC~2009-02-25-17:07:48 observation, where $\{\alpha, \zeta, \phi_{0}\} \sim \{13^{\circ}, 9^{\circ}, 20^{\circ}\}$, and the product of observed offset of the narrowband component from $\phi_{0}$ (i.e $\delta\phi \approx 90^{\circ}+ \phi_{0} \approx 110^{\circ}$), we obtain an angular offset of $\theta_{\mu\rightarrow\ell} \sim 16^{\circ}$.

\subsection{Magnetospheric plasma lens constraints}

The narrowband nature of the radio bursts from 1E~1547.0$-$5408, along with the double-peaked spectra detected among several examples shown in Figure~\ref{fig:bursts}, appear similar to the expected spectral features that arise from propagation through a plasma lens along the line of sight \cite{Cordes2017}.
In order for the radio pulses from the broadband component to remain unaffected, the lens must be located within the co-rotating magnetosphere of the magnetar and locked in phase with the narrowband component.
Given the lack of consistency in bandwidth and center frequency of subsequent bursts separated by $\lesssim 5$\,ms (see Extended Data Figure~\ref{fig:afb}), the plasma within the lens must be either highly dynamic or strongly restricted in scale.
We can estimate the scale of the lens as
\begin{equation}
    a = 2\pi \frac{\Delta t}{P} (R_{\rm lens}),
\end{equation}
where $\Delta t = 5$\,ms is the putative decorrelation timescale, $P = 2.069$\,s is the magnetar rotation period, and $R_{\rm lens}$ is the distance of the lens from the center of the magnetar.
An upper bound of $a \lesssim 1497$\,km can be derived assuming the lens is located at the light cylinder radius $R_{\rm LC} = 98591$\,km.
For a radio emission altitude of $\sim$300-1300\,km \cite{Stewart2025}, the distance between the source and a lens at the light cylinder is approximately equal to the lens altitude, i.e $d_{\rm ls} \approx R_{\rm lens}$.
From Ref. \cite{Cordes2017}, the gain of a 1D Gaussian lens can be computed as
\begin{equation}
    G = | 1 + \alpha_{\rm lens}(1 - 2u)e^{-u^{2}} |^{-1},
\end{equation}
where $u$ is the relative transverse offset of the lens along our line of sight to the radio source scaled by the lens scale, and $\alpha_{\rm lens}$ is a dimensionless parameter given by
\begin{equation}
    \alpha_{\rm lens} = \frac{3430\,{\rm DM_{lens}}\,d_{\rm sl}}{(\nu a)^{2}} \Bigg(\frac{d_{\rm lo}}{d_{\rm so}}\Bigg).
\end{equation}
Here, $\nu$ is the observing frequency in GHz, ${\rm DM_{lens}}$ is the dispersion measure of the plasma lens with units pc\,cm$^{-3}$, $a$ is the lens scale factor in AU, $d_{\rm sl}$ the distance between the source and the lens, $d_{\rm lo}$ the distance between the lens and the observer, and $d_{\rm so}$ the distance between the source and the observer. 
The latter three components have units of kpc.
In the case of a lens located at the light-cylinder radius of 1E~1547.0$-$5408, $d_{\rm lo} \approx d_{\rm so}$.
For the previously derived lens scale and $d_{\rm ls} \approx R_{\rm LC}$, we find that a dispersion measure of ${\rm DM_{lens}} \approx 1.5$\,pc\,cm$^{-3}$ is required to ensure the double caustic remains within the 8--9\,GHz band. 
A figure showing the Gain as a function of transverse offset and observing frequency for this scenario is provided in the Supplementary Materials.

A lens located within the magnetosphere of 1E~1547.0$-$540 would be comprised of a relativistic pair plasma.
The relativistic motion of the charged particles would suppress the plasma frequency, reducing the amount of dispersion imparted on any propagating radio waves.
Hence the charged particle density must be substantially higher within than the cold plasma regime in order to impart an equivalent amount of dispersion.
For a Gaussian lens, the dispersion measure can be estimated as
\begin{equation}
    {\rm DM_{\rm lens}} \approx \frac{n_{0} \sqrt{2\pi}\,a}{\langle \gamma \rangle},
\end{equation}
where $n_{0}$ is the charged particle density at the lens center and $\langle \gamma \rangle$ is the bulk Lorentz factor of the particles. 
In the above scenario of a plasma lens located at the light cylinder, the charged particle density would be $n_{0} \approx 1.2 \times 10^{10}\,{\rm cm}^{-3} \langle \gamma \rangle$.
For reference, the Goldriech-Julian plasma density for a neutron star with the same rotation period and spin-down derived magnetic field strength as 1E~1547.0$-$5408 is $n_{\rm GJ} = \frac{B}{P\,e\,c} \approx 7.4 \times 10^{12}$\,cm$^{-3}$. 
When scaled to the light cylinder radius, the plasma density drops to $n_{p} = n_{\rm GJ} (R_{\rm NS}/R_{\rm LC})^{3} \approx 7.6$\,cm$^{-3}$, approximately 10--12 orders of magnitude lower than required to sustain the plasma lens for bulk Lorentz factors between $\langle \gamma \rangle =1$--$100$.
Positioning the lens at a lower altitude would partially alleviate this tension, however there remains a strong dependence on the lens geometry and assumed validity of the lensing model in a hot relativistic pair plasma.

\clearpage

\begin{landscape}
\begin{table*}
\centering 
\caption{List of \textsl{Murriyang} observations of 1E~1547.0$-$5408, flux densities and the number of detected narrowband bursts. Entries delineated by a forward slash indicate where data was recorded by the DFB4 (left) and AFB (right). Values in parentheses indicate the 1-$\sigma$ uncertainties.}
\label{tab:obs}
\begin{tabular}{lccccccc}
\hline
\hline
Observation start & MJD & Frequency & Length & Sub-int. length & $S_{\rm mean}$ & No. narrowband & Burst rate  \\
(UTC) &  & (MHz) & (s) & (s) & (mJy) & bursts & (hr$^{-1}$) \\
\hline
2009-02-20-23:55:04 & 54882 & 3094 & 627 & 30 & 9.6(6) & 0 & 0 \\
2009-02-21-17:49:15 & 54883 & 3094 & 58 & 30 & 46.5(7) & 0 & 0 \\
2009-02-21-17:51:50 & 54883 & 3094 & 598 & 30 & 63.7(5) & 0 & 0 \\
2009-02-21-21:14:44 & 54883 & 3094 & 598 & 30 & 68.0(4) & 0 & 0 \\
2009-02-23-23:49:49/23:50:17 & 54885 & 8356/8452 & 567/594 & 30 & 51.1(5) & 2/3 & 13(4) \\
2009-02-24-14:56:43/14:57:22 & 54886 & 8356/8452 & 1769/1771 & 30 & 48.1(8) & 1/1 & 2(1) \\
2009-02-24-15:48:59/15:49:17 & 54886 & 8356/8452 & 1597/1573 & 20 & 57.3(4) & 2/2 & 5(2) \\
2009-02-25-17:07:48/17:08:12 & 54887 & 8356/8452 & 1058/1800 & 20 & 51.7(5) & 71/59 & 241(15) \\
2009-02-25-17:40:01/17:40:40 & 54887 & 8356/8452 & 277/285 & 20 & 48.7(4) & 8/4 & 104(10) \\
2009-02-26-14:16:00 & 54888 & 8356 & 97 & 20 & 22(3) & 0 & 0 \\
2009-02-26-14:20:08 & 54888 & 8356 & 1819 & 20 & 5.3(7) & 0 & 0 \\
2009-02-27-00:07:53 & 54889 & 8356 & 78 & 20 & 15.9(7) & 0 & 0 \\
2009-03-01-00:02:48 & 54891 & 8356 & 1118 & 20 & 11.9(7) & 0 & 0 \\ 
\hline\\
\end{tabular}
\end{table*}
\end{landscape}

\clearpage
\setcounter{figure}{0}

\begin{figure*}
    \centering
    \includegraphics[height=0.3\linewidth]{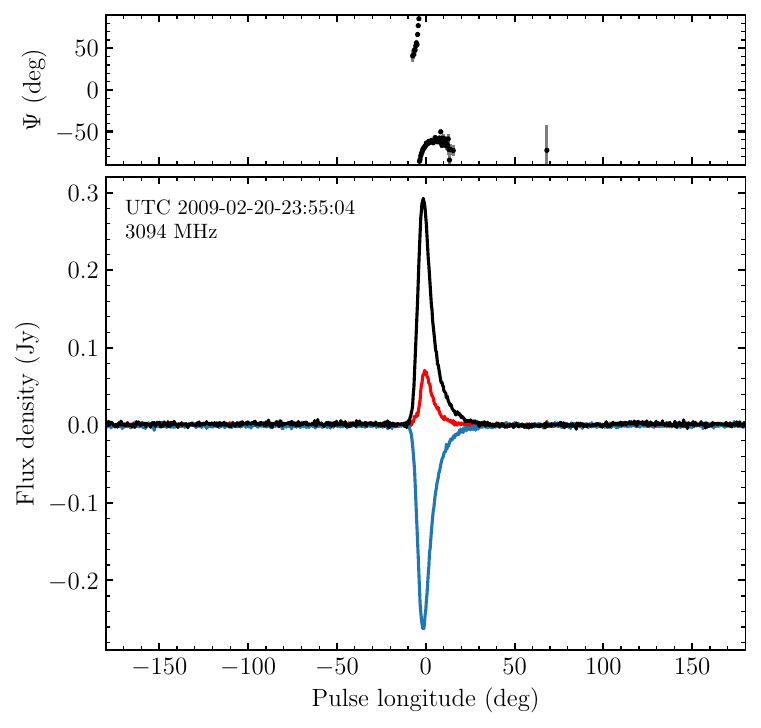}
    \includegraphics[height=0.3\linewidth]{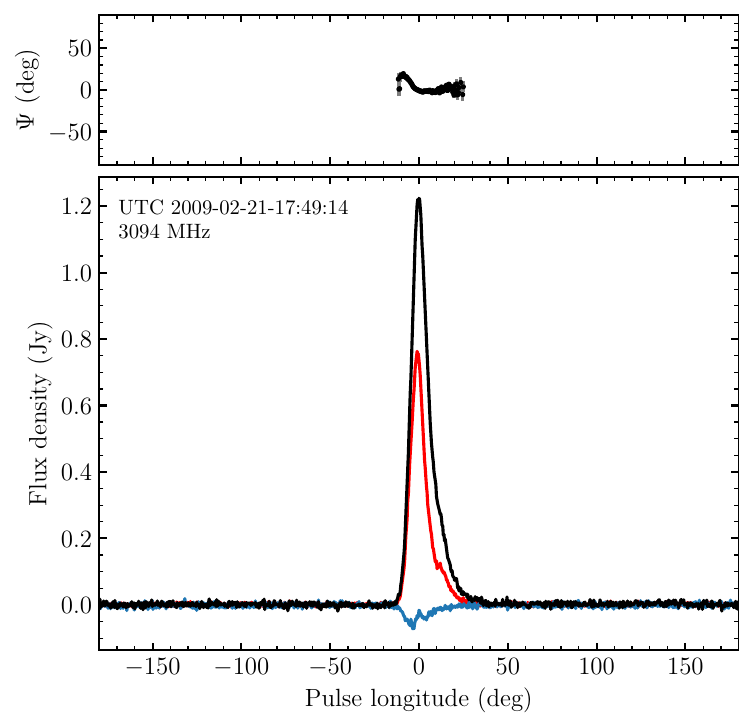}
    \includegraphics[height=0.3\linewidth]{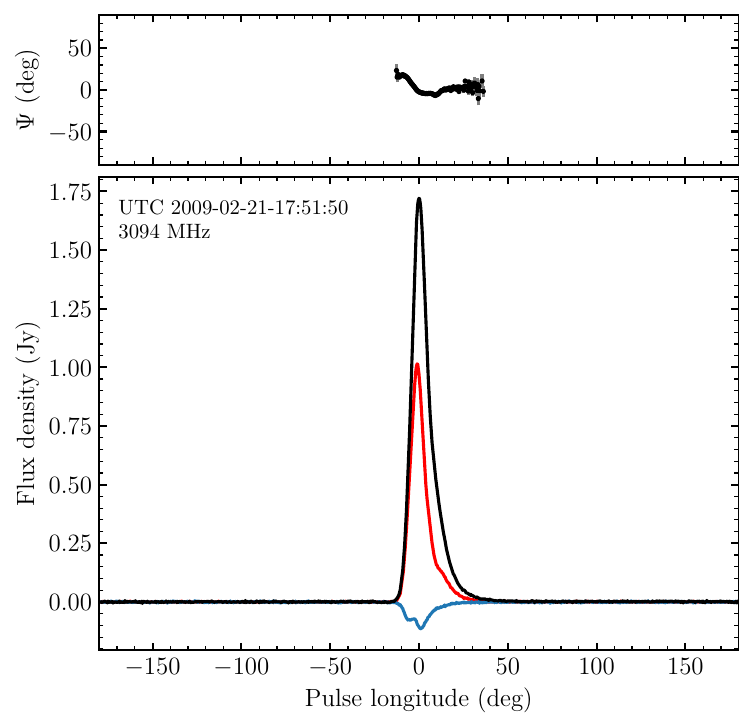}
    \includegraphics[height=0.3\linewidth]{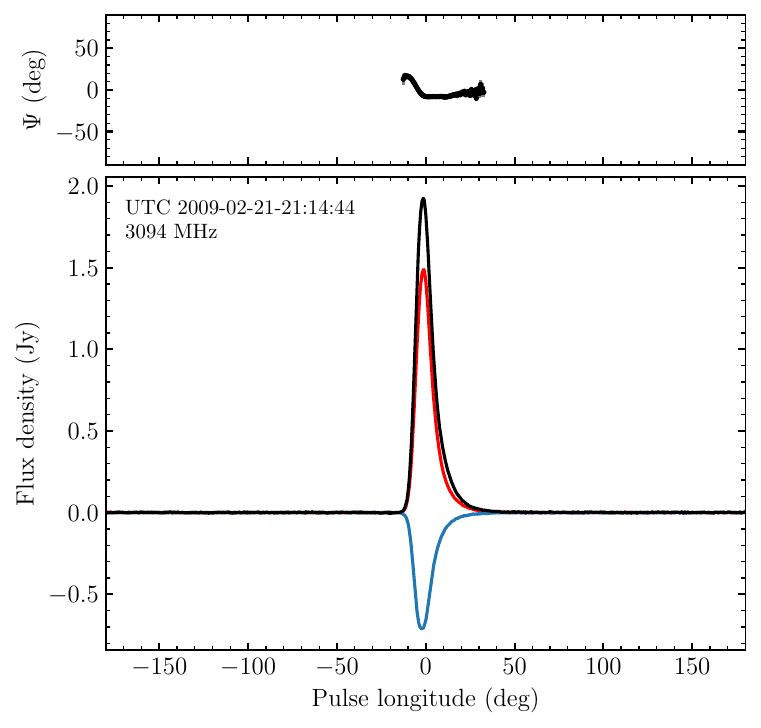}
    \includegraphics[height=0.3\linewidth]{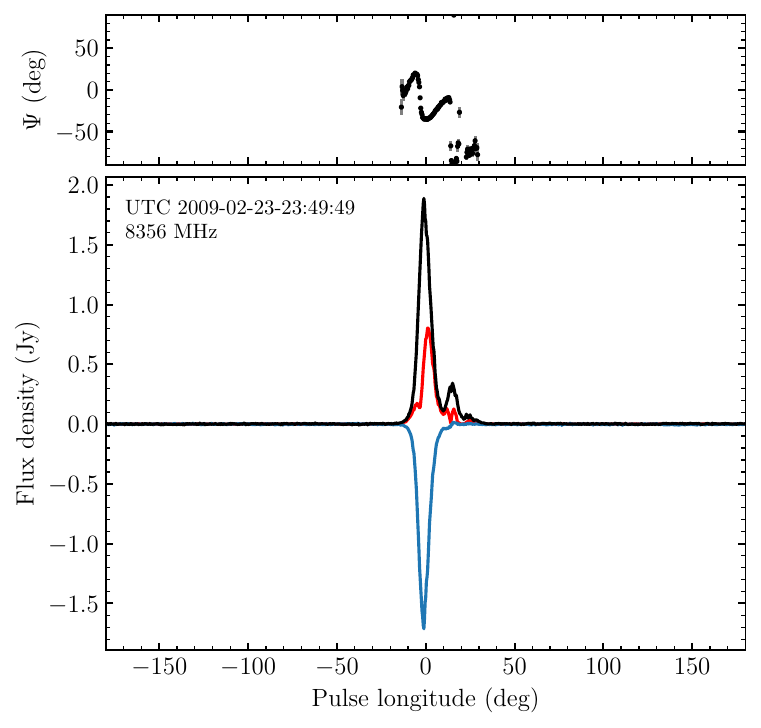}
    \includegraphics[height=0.3\linewidth]{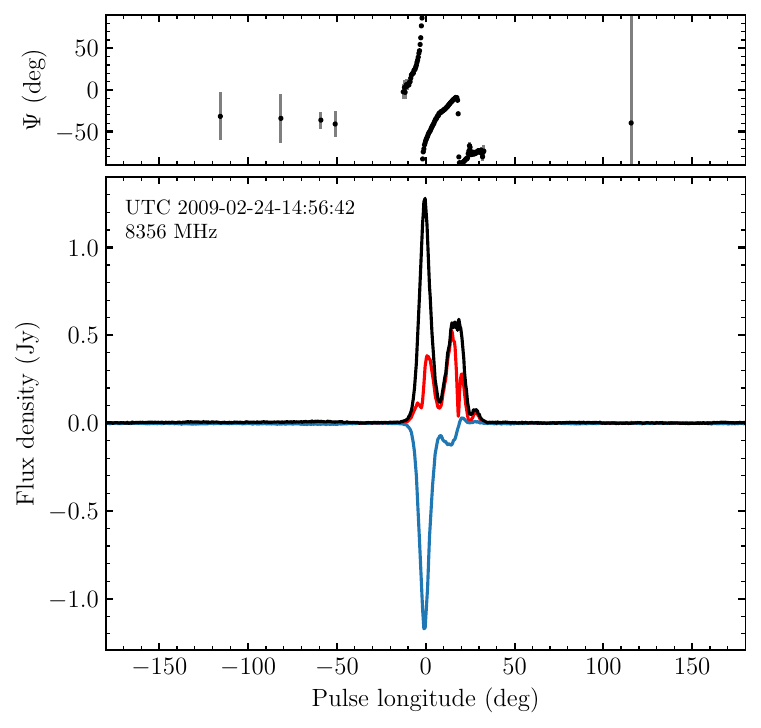}
    \includegraphics[height=0.3\linewidth]{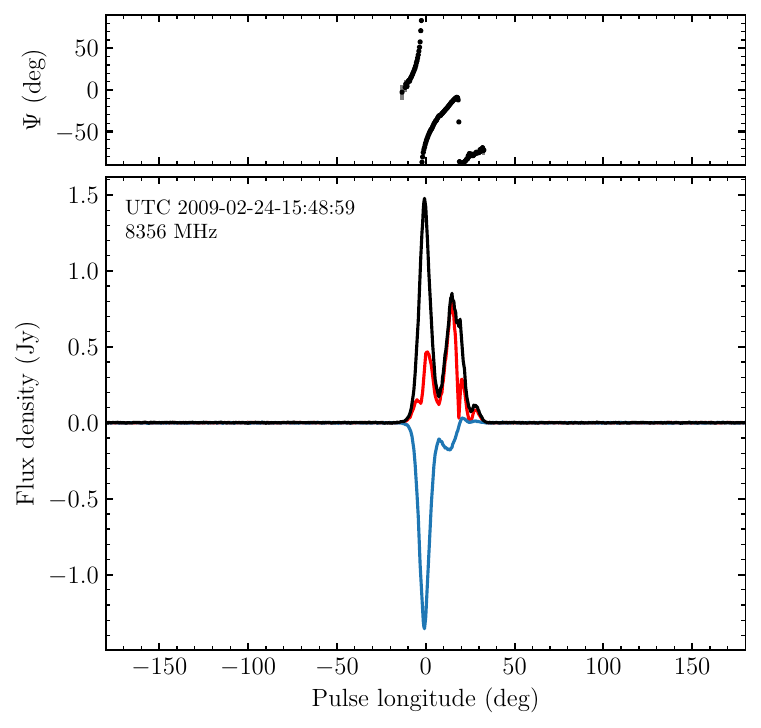}
    \includegraphics[height=0.3\linewidth]{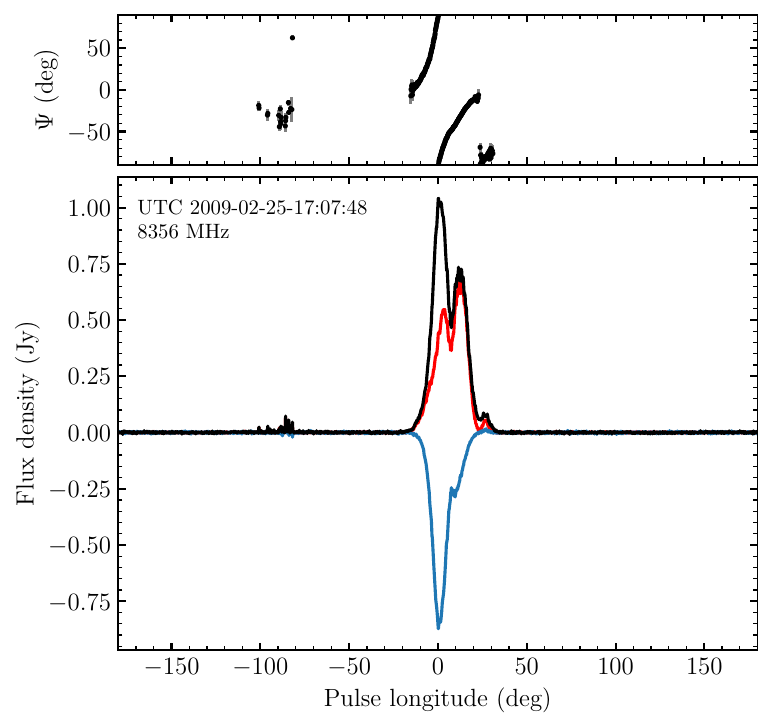}
    \includegraphics[height=0.3\linewidth]{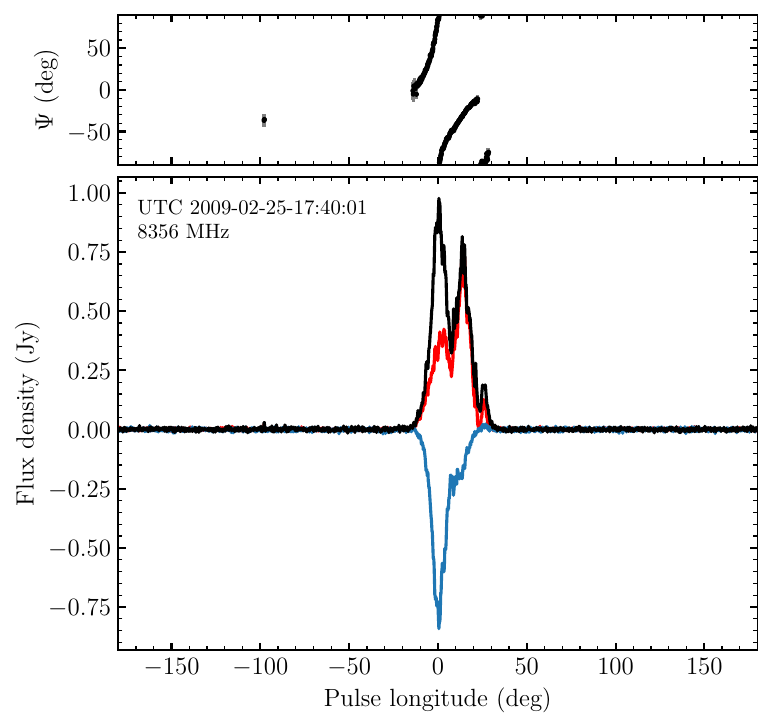}
    \includegraphics[height=0.3\linewidth]{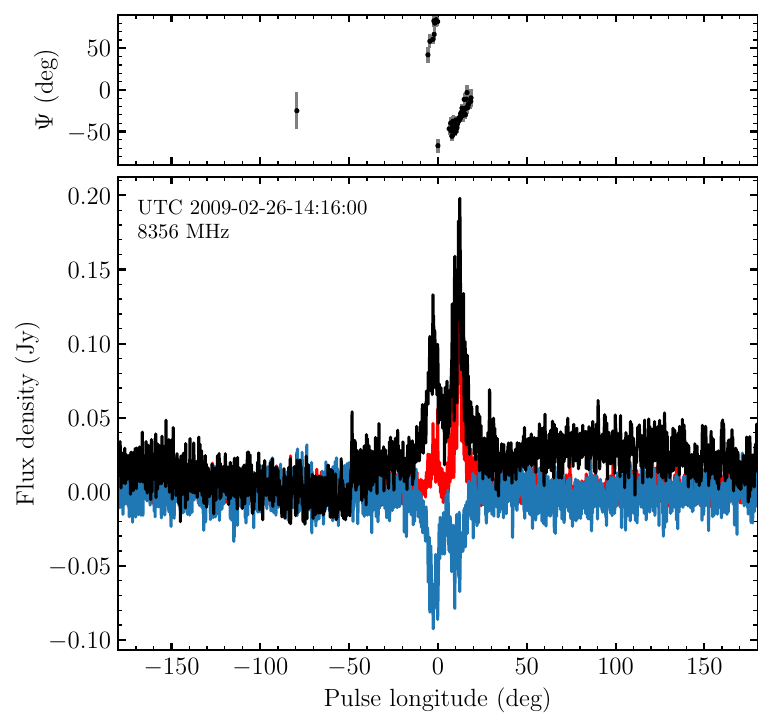}
    \includegraphics[height=0.3\linewidth]{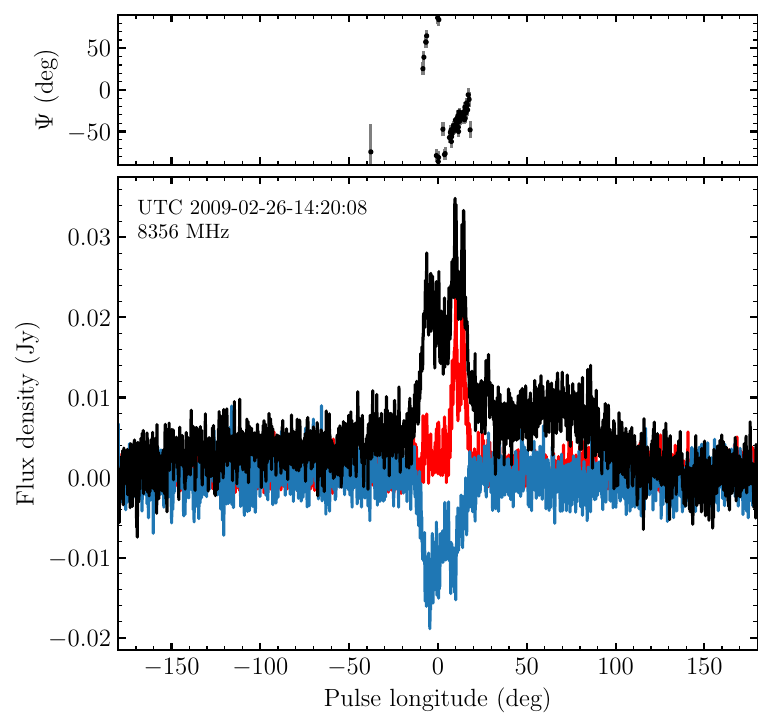}
    \includegraphics[height=0.3\linewidth]{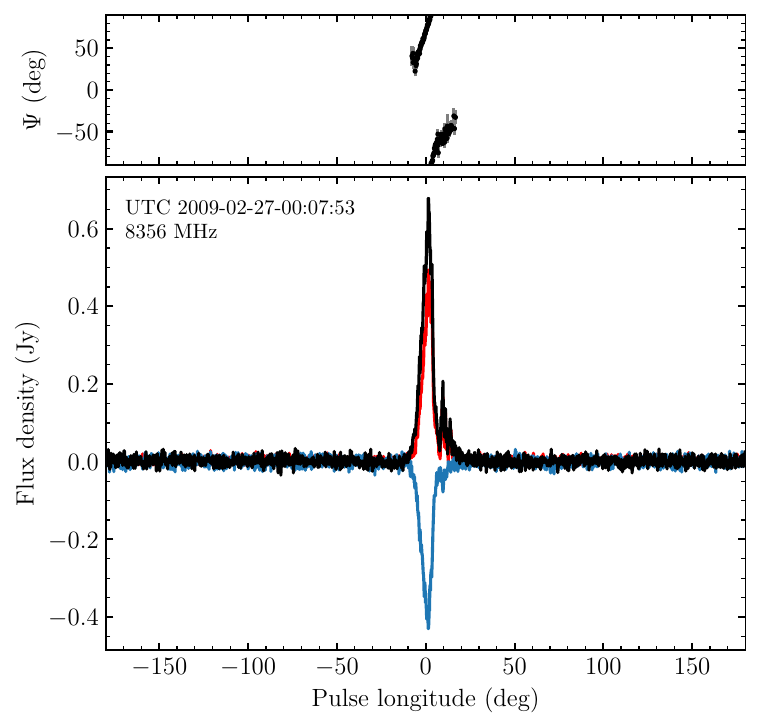}
    \includegraphics[height=0.3\linewidth]{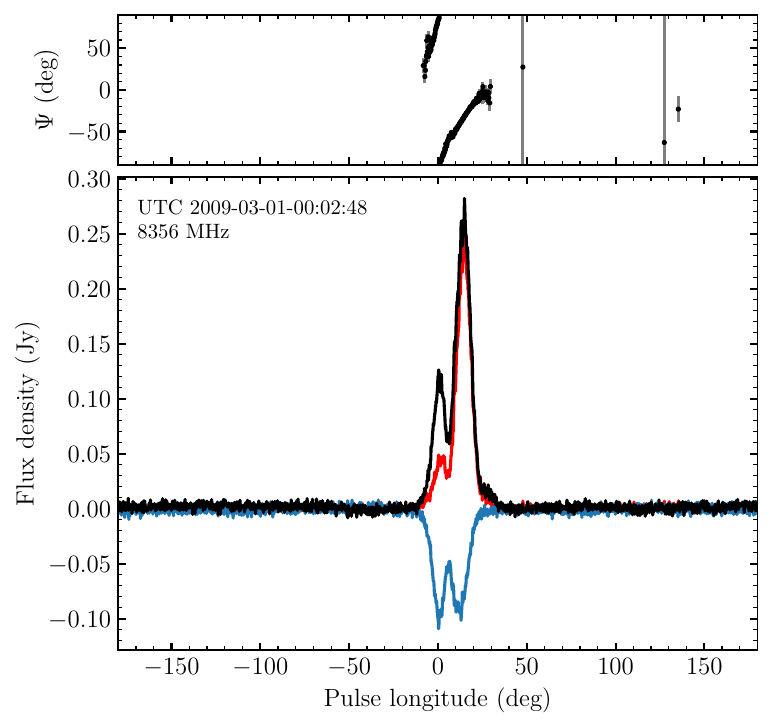}
    \caption{{\bf Time and frequency averaged polarization profiles of 1E~1547.0$-$5408 detected by Murriyang between 2009 February 20 to March 1.} The upper panel in each plot shows the linear polarization position angle ($\Psi$) after correcting for a Faraday rotation measure of $-1847.6$\,rad\,m$^{-2}$ at infinite frequency, and the lower panel the total intensity emission (black), total linear polarization (red) and circular polarization (blue).}
    \label{fig:profiles}
\end{figure*}

\clearpage

\begin{figure}
    \centering
    \includegraphics[width=0.8\linewidth]{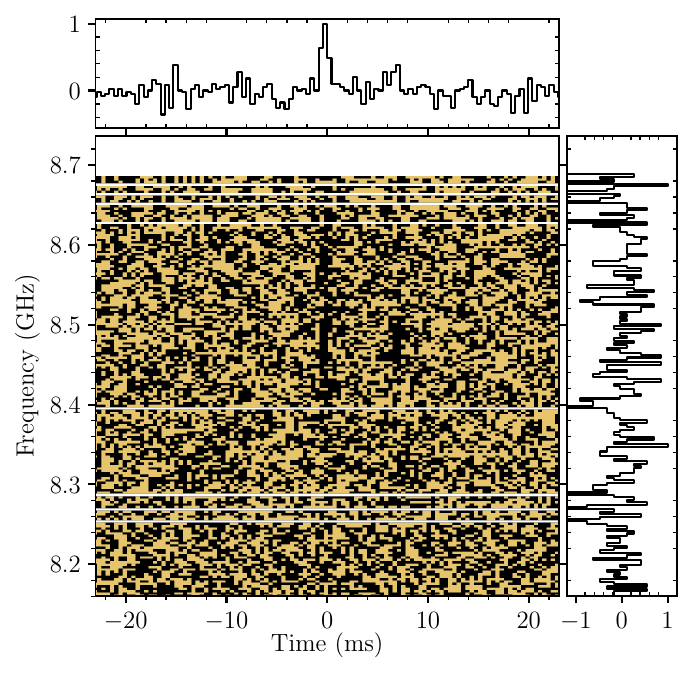}
    \caption{{\bf A pair of narrowband bursts recorded by the analogue filterbank signal processor that occurred within a single rotation of 1E~1547.0$-$5408.}  The fully frequency-averaged burst profile is shown in the top panel, while the dynamic spectrum of the bursts are shown in middle, and the time-averaged spectrum is presented in the right-hand panel.}
    \label{fig:afb}
\end{figure}

\clearpage

\begin{figure}
    \centering
    \includegraphics[width=0.8\linewidth]{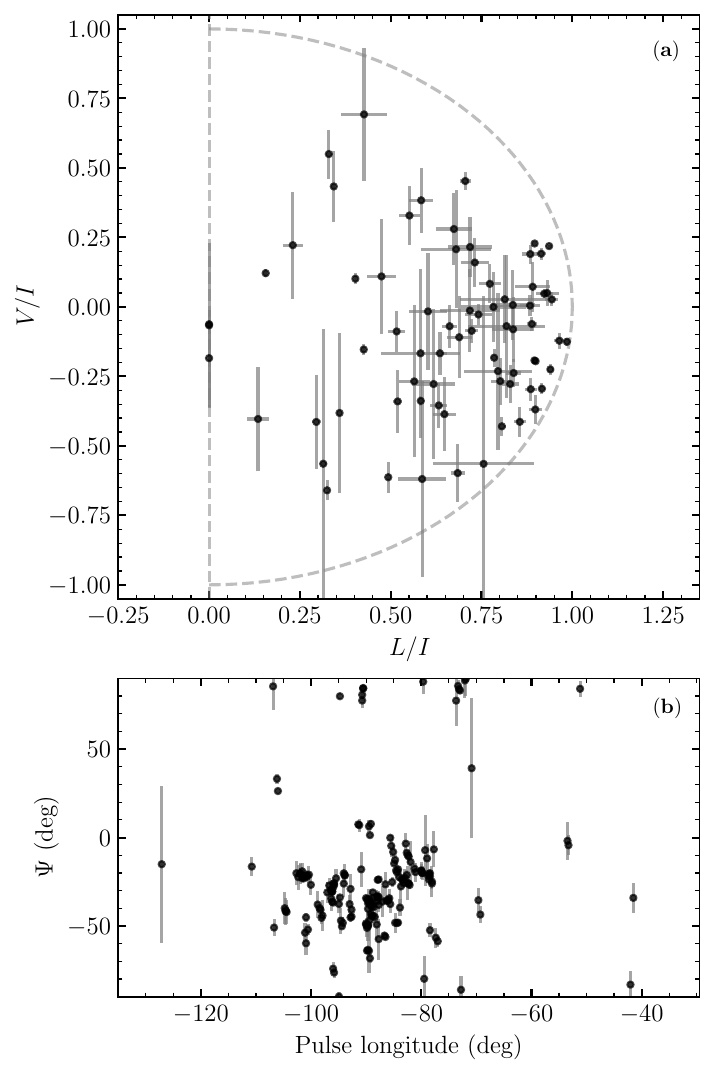}
    \caption{{\bf Polarization properties of the narrowband bursts.} Linear ($L/I$) and circular ($V/I$) polarization fractions are shown in ({\bf a}), where the dashed semi-circle indicates $\sqrt{L^{2} + V^{2}}/I = 1$. The distribution of linear PA swings ($\Psi$) across each burst are presented in ({\bf b}).}
    \label{fig:pol}
\end{figure}

\clearpage

\begin{figure}
    \centering
    \includegraphics[width=0.9\linewidth]{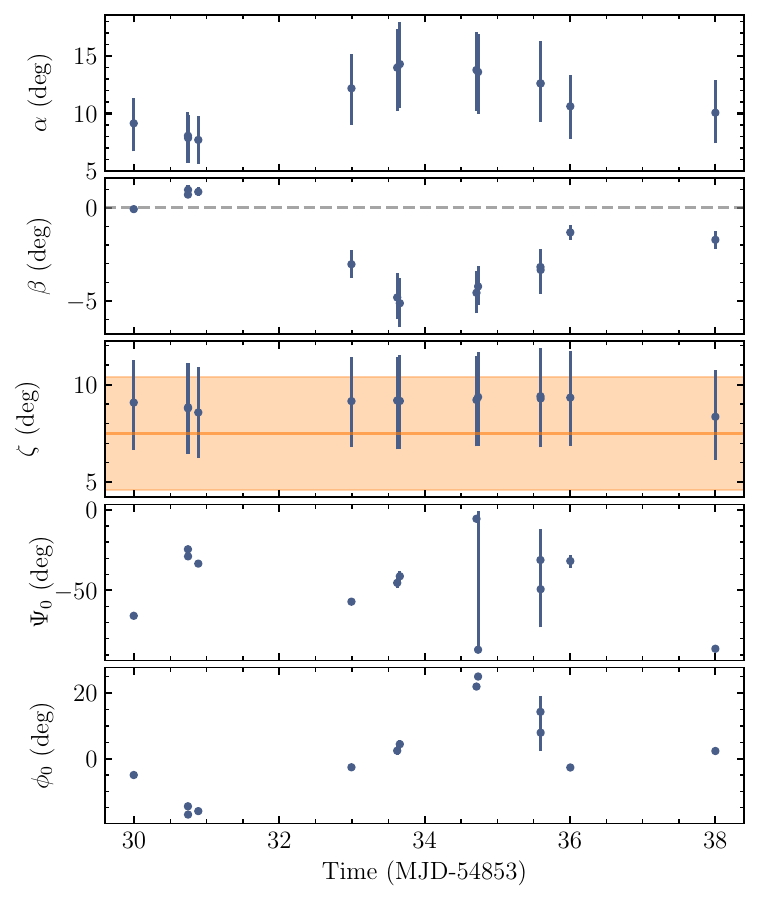}
    \caption{{\bf Changes in the recovered magnetic geometry of 1E~1547.0$-$5408 over our presented observing span.} We show the inclination angle ($\alpha$) in ({\bf a}), along with the angle between our line of sight and the magnetic ($\beta$) and rotation ($\zeta$) axes of the magnetar in ({\bf b}) and ({\bf c}). Panels ({\bf d}) and ({\bf e}) show the inflection point of the RVM ($\Psi_{0}$) and phase of the magnetic meridian  ($\phi_{0}$). The orange line and shading indicates the mean and 1-$\sigma$ width of the Gaussian prior on $\zeta$. Note the apparent sinusoidal changes with a $\sim 6.1$\,d period in $\alpha$, $\beta$, $\Psi_{0}$ and $\phi_{0}$.}
    \label{fig:rvm}
\end{figure}

\clearpage
\setcounter{figure}{0}

\begin{figure}
    \centering
    \includegraphics[width=\linewidth]{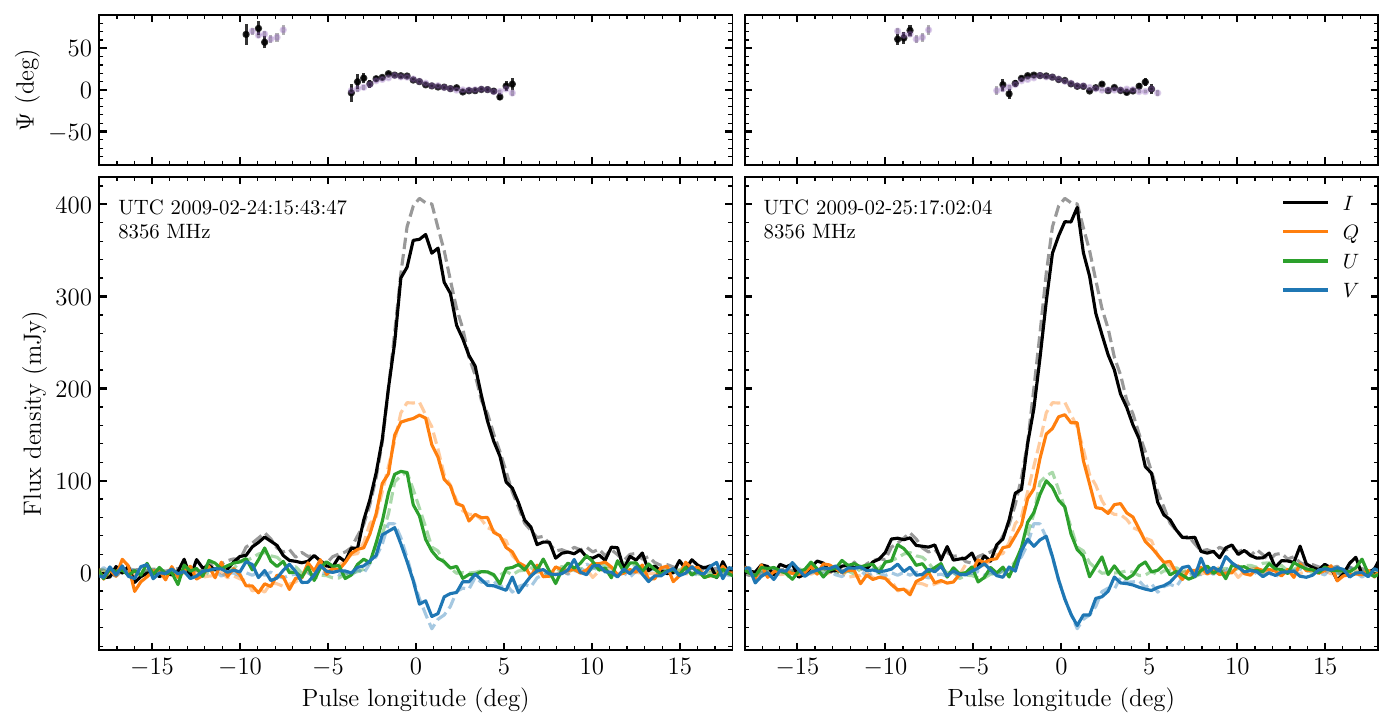}
    \caption{{\bf Comparison of the 8356\,MHz polarization profile of the reference pulsar PSR~B1641$-$45}. Panels show the linear polarization position angle swing ($\Psi$) (top), polarization profile with Stokes $I$ in black, Stokes $Q$ in orange, Stokes $U$ in green and Stokes $V$ in blue. The time-stamps indicates the UTC start time of the observation. The position angle of the reference observation from Ref.~\cite{Johnston2006} is shown in purple, while the corresponding Stokes parameters are presented as the dashed lines in the lower panels.}
    \label{fig:B1641}
\end{figure}

\begin{figure}
    \centering
    \includegraphics[width=\linewidth]{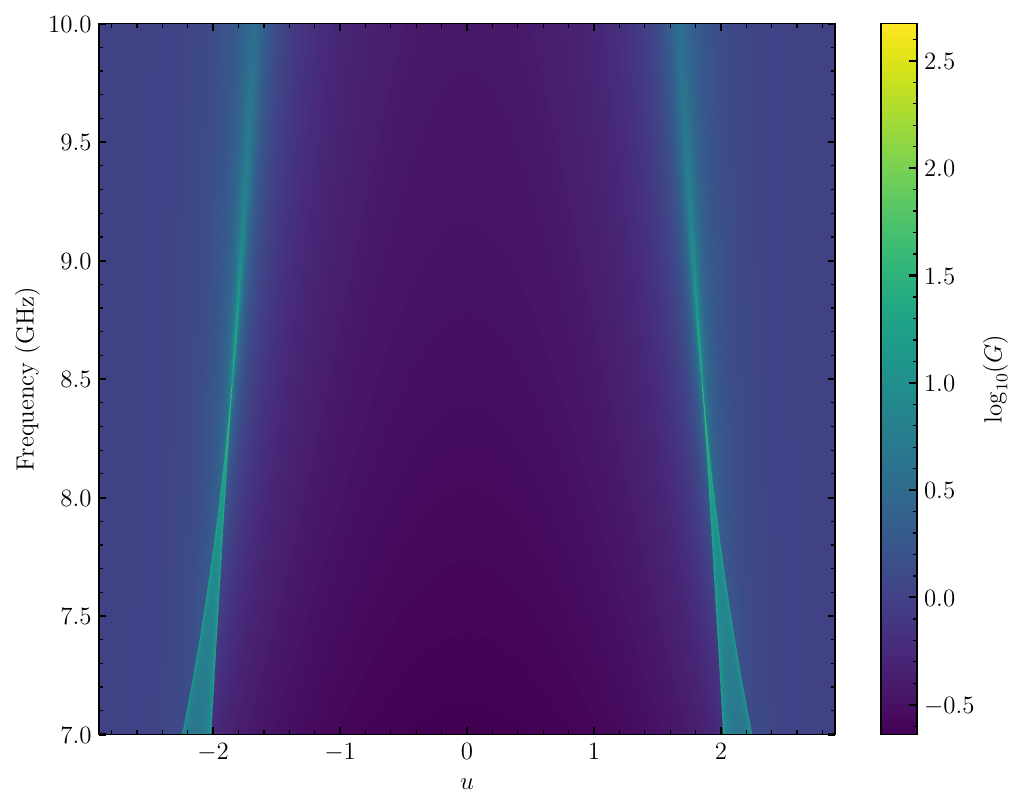}
    \caption{{\bf Simulated gain of a plasma lens as a function of transverse offset and observing frequency}. The lens is assumed to be placed at the light-cylinder radius of the magnetar, where $d_{\rm ls} = R_{\rm LC} = 98591$\,km, with a lens scale of $a = 1497$\,km and a dispersion measure of ${\rm DM_{ lens}} = 1.5$\,pc\,cm$^{-3}$.}
    \label{fig:B1641}
\end{figure}

\clearpage

\section*{Declarations}

\bmhead{Acknowledgments}

We thank Zorawar Wadiasingh and Matthew Baring for stimulating discussions on the X-ray and radio emission regions of magnetar magnetospheres.
Murriyang, the CSIRO’s Parkes radio telescope, is part of the Australia Telescope National Facility (\href{https://ror.org/05qajvd42}{https://ror.org/05qajvd42}) which is funded by the Australian Government for operation as a National Facility managed by CSIRO.
We acknowledge the Wiradjuri people as the traditional owners of the Observatory site.
This project was supported by resources and expertise provided by CSIRO IMT Scientific Computing, and made use of the Ngarrgu Tindebeek supercomputer at the OzSTAR national facility at Swinburne University of Technology. 
The OzSTAR program receives funding in part from the Astronomy National Collaborative Research Infrastructure Strategy (NCRIS) allocation provided by the Australian Government, and from the Victorian Higher Education State Investment Fund (VHESIF) provided by the Victorian Government.
This work made use of NASA's Astrophysics Data System.
MEL is supported by an Australian Research Council (ARC) Discovery Early Career Research Award (DE250100508).
GY work is supported by NASA under award number 80GSFC24M0006.

\bmhead{Conflicts of interest}
The authors declare no competing interests. 

\bmhead{Ethics approval and consent to participate}
Not applicable.
\bmhead{Consent of publication}
Not applicable.
\bmhead{Data availability}
The raw Parkes data is available to download via the CSIRO Data Access Portal (\href{https://data.csiro.au/}{https://data.csiro.au/} -- see Refs. \cite{csiro:P456-2008OCTS, csiro:P602-2008OCTS-01, csiro:P602-2008OCTS, csiro:PUNDEF-2009}).
Other data products are available upon reasonable request to the corresponding author.
\bmhead{Materials availability}
Not applicable.
\bmhead{Code availability}
Data reduction and analysis of radio data was performed using {\tt dspsr} (\url{https://dspsr.sourceforge.net/}), {\tt psrchive} (\url{https://psrchive.sourceforge.net/}) and {\tt bilby} (\url{https://github.com/bilby-dev/bilby}). Timing analysis was performed using {\tt tempo2} (\url{https://github.com/mattpitkin/tempo2}) and {\tt PINT} (\url{https://github.com/nanograv/PINT}). Custom codes for making the X-ray profiles and specific radio analyses are available upon reasonable request from corresponding authors.  

\bmhead{Author's Contributions}
M.E.L. performed the radio data reduction and analysis, and led the writing of the paper.
P.S. conducted the X-ray data reduction and contributed to writing the paper.
F.C. was the PI of the Parkes P602 project under which most of these data were collected, and contributed to writing of the paper.
D.M.P. led the search for X-ray/gamma-ray bursts in the Swift Burst Alert Telescope data and contributed to writing the paper.
J.E.R. and J.M.S. conducted the Parkes observations.
L.J.T. was responsible for recovering the analogue filterbank data and converting it into a format that could interface with modern software tools.
G.Y. led the search for X-ray bursts in archival Fermi Gamma-ray Burst Monitor data and contributed to writing the paper.

\end{document}